\input harvmac.tex
\input epsf
\noblackbox

\newcount\figno
 \figno=0
 \def\fig#1#2#3{
\par\begingroup\parindent=0pt\leftskip=1cm\rightskip=1cm\parindent=0pt
 \baselineskip=11pt
 \global\advance\figno by 1
 \midinsert
 \epsfxsize=#3
 \centerline{\epsfbox{#2}}
 \vskip 12pt
 {\bf Fig.\ \the\figno: } #1\par
 \endinsert\endgroup\par
 }
 \def\figlabel#1{\xdef#1{\the\figno}}

\newdimen\tableauside\tableauside=1.0ex
\newdimen\tableaurule\tableaurule=0.4pt
\newdimen\tableaustep
\def\phantomhrule#1{\hbox{\vbox to0pt{\hrule height\tableaurule width#1\vss}}}
\def\phantomvrule#1{\vbox{\hbox to0pt{\vrule width\tableaurule height#1\hss}}}
\def\sqr{\vbox{%
  \phantomhrule\tableaustep
  \hbox{\phantomvrule\tableaustep\kern\tableaustep\phantomvrule\tableaustep}%
  \hbox{\vbox{\phantomhrule\tableauside}\kern-\tableaurule}}}
\def\squares#1{\hbox{\count0=#1\noindent\loop\sqr
  \advance\count0 by-1 \ifnum\count0>0\repeat}}
\def\tableau#1{\vcenter{\offinterlineskip
  \tableaustep=\tableauside\advance\tableaustep by-\tableaurule
  \kern\normallineskip\hbox
    {\kern\normallineskip\vbox
      {\gettableau#1 0 }%
     \kern\normallineskip\kern\tableaurule}%
  \kern\normallineskip\kern\tableaurule}}
\def\gettableau#1 {\ifnum#1=0\let\next=\null\else
  \squares{#1}\let\next=\gettableau\fi\next}

\tableauside=1.0ex \tableaurule=0.4pt
\def\a{{\alpha}}

\def\R{{{\cal R}}}

\def\z{{\zeta}}

\def\e{{\epsilon}}
\def\wo{{\omega}}
\def\vert{{|}}

\def\k{{ k }}

\def\frac#1#2{{#1\over #2}}
\def\half{{ {\frac{1}{2}} }}

\lref\osv{H.~Ooguri, A.~Strominger, and C.~Vafa,
``Black hole attractors and the topological string,''
Phys.\ Rev.\ D70 (2004) 106007, {\tt hep-th/0405146}.
}
\lref\V{C.~Vafa, ``Two-dimensional Yang-Mills, black holes and topological
strings,'' {\tt hep-th/0406058}.}
\lref\gt{D.~Gross and W.~Taylor, ``Two-dimensional QCD is a string theory,''
Nucl.\ Phys.\ B400 (1993) 181-210, {\tt  hep-th/9301068};
``Twists and Wilson loops in the string theory of two-dimensional QCD,''
Nucl.\ Phys.\ B403 (1993) 395-452, {\tt hep-th/9303046}.}
\lref\BP{J.~Bryan and R.~Pandharipande, ``Local Gromov-Witten
theory of curves,'' {\tt math.AG/0411037}.}
\lref\AMViii{
M.~Aganagic, A.~Klemm, M.~Marino, and C.~Vafa,
``Matrix model as a mirror of Chern-Simons theory,''
JHEP\ 0402 (2004) 010,
{\tt hep-th/0211098}.
}
\lref\W{E.~Witten, ``Two-dimensional gauge theories revisited,''
J.\ Geom.\ Phys.\ 9 (1992) 303, {\tt hep-th/9204083}.}
\lref\BT{M.~Blau and  G.~Thompson, ``Derivation of the Verlinde
formula from Chern-Simons theory and the $G/G$ model,''
Nucl.\ Phys.\ B408 (1993) 345, {\tt hep-th/9305010}.}
\lref\JP{E.~Witten, ``Quantum field theory and the Jones polynomial,'' Com.\
Math.\ Phys. 121 (1989) 351.}

\lref\VW{C.~Vafa and E.~Witten, ``A strong coupling test of
S-duality,'' Nucl.\ Phys.\ B 431 (1994) 3, {\tt  hep-th/9408074}.}
\lref\Rudd{R.~Rudd, ``String partition function for QCD on the
torus,'' {\tt hep-th/9407176}.}

\lref\Klim{C.Klimcik, ``The formulae of Kontsevich and Verlinde from
the perspective of the Drinfeld double,'' Commun. Math. Phys. 217
(2001) 203-228, {\tt hep-th/9911239}.}

 \lref\KS{V.~Kazakov,
M.~Staudacher, and T.~Wynter, ``Advances in large $N$ group theory
and the solution of two-dimensional $R^2$ gravity'', Proceedings of
Nato Advanced Summer Institute {\it Low Dimensional Applications of
Quantum Field Theory} (Cargese, 1995), {\tt hep-th/9601153}.}
\lref\ghv{D.~Ghoshal and C.~Vafa, ``$c = 1$ string as the
topological theory of the conifold,'' Nucl.\ Phys.\ B 453 (1995)
121, {\tt hep-th/9506122}.
}
\lref\IKii{A.~Iqbal and A.~Kashani-Poor, ``The vertex on a strip,''
{\tt hep-th/040174}.}
\lref\AKMV{
M.~Aganagic, A.~Klemm, M.~Marino, and C.~Vafa,
``The topological vertex,''
{\tt hep-th/0305132}.
}
\lref\KazakovPM{
V.~Kazakov, I.~K.~Kostov and D.~Kutasov,
``A matrix model for the two-dimensional black hole,''
Nucl.\ Phys.\ B { 622} (2002) 141,
{\tt hep-th/0101011}.
}
\lref\AMV{
M.~Aganagic, M.~Marino, and C.~Vafa,
``All loop topological string amplitudes from Chern-Simons theory,''
Commun.\ Math.\ Phys.\  247 (2004) 467 (2004),
{\tt hep-th/0206164}.
}
\lref\ovc{H.~Ooguri and C.~Vafa,
``Two-dimensional black hole and singularities of CY manifolds,''
Nucl.Phys. B463 (1996) 55-72, {\tt hep-th/9511164}.
}
\lref\roche{
E.~Buffenoir and P.~Roche,
``Two-dimensional lattice gauge theory based on a quantum group,''
Commun.\ Math.\ Phys.\  {\bf 170}, 669 (1995)
[arXiv:hep-th/9405126].
}
\lref\INOV{A.~Iqbal, N.~Nekrasov, A.~Okounkov, and
C.~Vafa, ``Quantum foam and topological strings",
{\tt hep-th/0312022}.}
\lref\OPN{D.~Maulik, N.~Nekrasov, A.~Okounkov, and R.~Pandharipande,
``Gromov-Witten theory and Donaldson-Thomas theory,"
math.AG/0312059 }
\lref\NV{A.~Neitzke and C.~Vafa,
     ``${\cal N}=2$ strings and the twistorial Calabi-Yau," {\tt hep-th/0402128}.}
\lref\NOV{N. Nekrasov, H. Ooguri, C. Vafa, ``S-duality and topological
strings," JHEP\ 0410 (2004) 009, {\tt hep-th/0403167}.
}
\lref\ADKMV{
M.~Aganagic, R.~Dijkgraaf, A.~Klemm, M.~Marino and C.~Vafa,
``Topological strings and integrable hierarchies,''
{\tt hep-th/0312085}.
}
\lref\wd{H. Ooguri and C. Vafa, ``Worldsheet
derivation of large $N$ duality,'' Nucl.\ Phys.\ B641 (2002) 3,
{\tt hep-th/0205297}.
}
\lref\OOV{H. Ooguri and C. Vafa, ``Knot invariants and topological
strings," Nucl.\ Phys.\ B577 (2000) 419-438, {\tt hep-th/991213}.
}
\lref\yoshioka{K.~Yoshioka,
``Betti numbers of moduli of stable sheaves on some surfaces,''
Proceedings of ICTP Conference {\it S-Duality and Mirror
Symmetry} (Trieste, 1995),
Nucl.\ Phys.\ B\ Proc.\ Suppl.\ 46 (1996) 263.}
\lref\nakajima{H.~Nakajima, ``Instantons on ALE spaces, quiver varieties
and Kac-Moody algebras,''
Duke\ Math.\ J.\ 76 (1994) 365.}
\lref\Nekrasov{N.~Nekrasov,
``Seiberg-Witten prepotential from instanton counting,''
Adv.\ Theor.\ Math.\ Phys.\  7 (2004) 831, {\tt hep-th/0206161}.
}
\lref\DV{R.~Dijkgraaf and C.~Vafa, ``Matrix models, topological strings,
and supersymmetric gauge theories'',
  Nucl.\ Phys.\ B644 (2002) 3, {\tt hep-th/0206255}. }
\lref\MM{
M.~Aganagic, A.~Klemm, M.~Marino and C.~Vafa,
``Matrix model as a mirror of Chern-Simons theory,''
JHEP {\bf 0402}, 010 (2004)
[arXiv:hep-th/0211098].
}
\lref\IK{A.~Iqbal and A.~Kashani-Poor,
``Instanton counting and Chern-Simons theory,''
Adv.\ Theor.\ Math.\ Phys.\ 7 (2003) 459-499, {\tt hep-th/0212279}.}
\lref\DMP{R.~Dijkgraaf, G.~Moore, R.~Plesser, ``The partition
function of 2d string theory,'' Nucl.\ Phys.\ B394 (1993) 356,
{\tt hep-th/9208031}.}

\Title{
  \vbox{\baselineskip12pt \hbox{hep-th/0411280}
  \hbox{CALT-68-2529}
\hbox{HUTP-04/A0049}
\hbox{UCB-PTH/04/33}
  \vskip-.5in}
}{\vbox{
  \centerline{Black Holes, $q$-Deformed 2d Yang-Mills,}
\smallskip
\centerline{and}
\smallskip
  \centerline{ Non-perturbative Topological Strings}
}}

\centerline{Mina Aganagic,$^{1}$ Hirosi Ooguri,$^2$ Natalia Saulina,$^3$
and Cumrun Vafa$^3$}
\bigskip\medskip
\centerline{$^1$ \it University of California, Berkeley, CA 94720, USA}
\vskip .03in
\centerline{$^2$ \it California Institute of Technology, Pasadena,
CA 91125, USA}
\vskip .03in
\centerline{$^3$ \it Jefferson Physical Laboratory,
Harvard University, Cambridge, MA 02138, USA}
\medskip
\medskip
\medskip
\vskip .5in


We count the number of bound states of BPS black holes on
local Calabi-Yau three-folds
involving a Riemann surface of genus $g$.  We show that the corresponding
gauge theory on the brane reduces to a $q$-deformed Yang-Mills theory
on the Riemann surface. Following the recent connection between the
black hole entropy and the
topological string partition function, we find that for a large black hole
charge $N$, up to corrections of $O(e^{-N})$, $Z_{BH}$ is given as
a sum of a square of chiral blocks, each of which corresponds
to a specific D-brane amplitude.
The leading chiral block, the vacuum
block, corresponds to the closed topological
string amplitudes. The sub-leading chiral blocks involve
topological string amplitudes with D-brane insertions
at $(2g-2)$ points on the Riemann surface analogous to
the $\Omega$ points in the large $N$ 2d Yang-Mills theory.
The finite $N$ amplitude provides a non-perturbative
definition of topological strings in these backgrounds.  This
also leads to a novel non-perturbative formulation of $c=1$ non-critical
string at the self-dual radius.

\Date{November, 2004}
\newsec{Introduction}

Counting of 4-dimensional BPS black hole microstates
arising upon compactifications
of type II superstrings on Calabi-Yau 3-folds has been recently
connected to topological string amplitudes in a highly non-trivial
way \osv.  In particular it has been argued that
for a large black hole charge $N$ and to all order in $1/N$ expansion,
the mixed ensemble partition function $Z_{BH}$
of BPS black holes is related to topological
string amplitudes $Z_{{\rm top}}$:
\eqn\conj{Z_{BH}=|Z_{{\rm top}}|^2.}
In fact the proposal in \osv\ goes further and uses the above relation
as defining what one means by the non-perturbative topological string
amplitude, including $O(e^{-N})$ corrections.

One main obstacle in checking this relation is that both sides
are difficult (at present impossible) to compute for compact
Calabi-Yau manifolds.  However, the logic of \osv\ can be adapted
to non-compact Calabi-Yau manifolds and the counting of BPS states
in these geometries.  In that context one could hope to check
this statement.

A first case where this was actually done \V\
was in the case of a local Calabi-Yau involving the sum of two
line bundles over a torus $T^2$.  In that case, not only it
was shown that \conj\ is correct to all orders in the $1/N$ expansion,
but that, at finite $N$, the notion of a holomorphic $Z_{top}$ ceases to make
sense due to $O(e^{-N})$ effects.  In fact counting
of BPS bound states in that case reduces to computing the partition
function of the 2d Yang-Mills theory on $T^2$, whose large $N$
expansion has the above chiral decomposition valid only to all orders
in the $1/N$ expansion, as was demonstrated in \gt.

The aim of this paper is to extend the computation in \V\ to the
case of a Calabi-Yau manifold where $T^2$ is replaced by
an arbitrary genus $g$ Riemann surface. Once again we find that
the topological gauge
theory on the brane reduces to 2d Yang-Mills theory on the Riemann
surface, with one additional subtlety:  the Yang-Mills theory
is {\it q-deformed}\ !  We then ask if the relation \conj\ holds
in this case.  Luckily the topological string amplitude for this
geometry has been computed very recently \BP ,
and so the right hand side of \conj\ is also known.  In checking
this relation we find an interesting new subtlety:  We find that
the relation \conj\ to all orders in the $1/N$
expansion is modified in this case to
\eqn\nec{Z_{BH}=\int \prod_{i=1}^{|2g-2|} dU_i\;\;
|Z_{{\rm top}}(U_1,...,U_{|2g-2|})|^2,}
where $Z_{{\rm top}}(U_1,...,U_{|2g-2|})$ is the topological string
amplitude with $|2g-2|$ stacks of D-branes inserted,
and $U_i$ corresponds to the
holonomy of the gauge field turned on in the $i$-th stack of D-branes.
The leading piece of this expansion is the closed string amplitude.
To extract this piece, the integral over the unitary group in the
above can be performed and the relation can be recast as
$$Z_{BH}=\sum_{R_1,...,R_{|2g-2|}}\; |Z^{{\rm top}}_{R_1,...,R_{|2g-2|}}|^2,$$
where
$$Z_{{\rm top}}(U_1,...,U_{|2g-2|})
=\sum_{R_1,...,R_{|2g-2|}}
\; Z^{{\rm top}}_{R_1,...,R_{|2g-2|}}\; \Tr_{R_1} U_1 \cdots
\Tr_{R_{|2g-2|}}U_{|2g-2|},$$
and $R_1,...,R_{|2g-2|}$ are representations of $SU(\infty)$ and
$\Tr_{R_i}U_i$ ($i=1,...,|2g-2|$) are their characters.
The closed string amplitude is the vacuum chiral block given
by $R_i=0$.

The organization of this paper is as follows:  In section 2 we review
perturbative results for topological strings on a Calabi-Yau with
a local geometry involving a Riemann surface.  We also present
a short alternative derivation of the main result.  In section 3 we study the
gauge theory living on a BPS brane in this geometry and
compute its partition function.
In particular we show that the relevant
field theory on the brane gets mapped to the $q$-deformed 2d Yang-Mills
theory on the Riemann surface.  In section 4 we use this result to count
the BPS black
hole degeneracy in this geometry.
In section 5 we take the limit of large black hole charges
and relate our result to the expected topological string amplitudes
reviewed in section 2, in accordance with \osv .
In section 6 we  discuss the limit of small area for a local ${\bf P}^1$
geometry (the resolved conifold). This is related by the mirror symmetry
to the deformed conifold and thus to the
$c=1$ non-critical bosonic string at the self-dual radius.  There we find
that the non-perturbative formulation involves chiral blocks which
represent a condensation of
a coherent state of tachyons
which emits tachyons at all frequencies.
In appendices A - D we collect some identities needed in the paper.

\newsec{Perturbative Topological String Theory}

In a recent work \BP\
topological A-model string theory amplitudes
$$
X= L_1\oplus L_2 \rightarrow \Sigma,
$$
were computed to all orders of perturbation theory in terms of a
certain topological quantum field theory on the Riemann surface
$\Sigma$. The total space $X$ is Calabi-Yau when the Chern class
of the normal bundle to $\Sigma$ cancels the canonical class, $i.e.$
$$
{\rm deg}(L_1)+{\rm deg}(L_2) = - \chi(\Sigma).
$$
For example, when $\Sigma$ is a closed Riemann surface of genus $g$,
we can take $({\rm deg}(L_1), {\rm deg}(L_2))=(p+2 g-2, -p)$ for any integer $p$.

In \BP\ a Riemann surface $\Sigma$ is viewed as obtained by
gluing of more basic building blocks: pants (P), annuli (A) and caps
(C).  The whole Calabi-Yau $X$ arises in this way from gluing simpler
3-folds which are rank $2$ holomorphic bundles over Riemann surfaces
with boundaries.  Note that under gluing, the Euler characters of the base
curves add, but moreover, the Chern classes of the normal bundles add.

Consider a Calabi-Yau $X$ involving a Riemann surface
$\Sigma_{g,h}$ of genus $g$ with $h$ punctures or equivalently $h$
semi-infinite cylinders. The holomorphic maps from a worldsheet to
$X$ are necessarily holomorphic maps to $\Sigma_{g,h}$. To get
non-trivial such maps of finite area, we must add Lagrangian
D-branes which circle the punctures (and that are $2$-dimensional
in the fiber direction) and consider holomorphic maps with
boundaries on the D-branes. In this way, we can view cutting and
pasting of the base Riemann surfaces and the corresponding
Calabi-Yau's as either adding or cancelling off D-branes. The
operations of gluing manifolds lead to composition of topological
A-model amplitudes that satisfy all axioms of a two dimensional
topological quantum field theory \BP . In the course of this
paper, we will explain from the physical standpoint what the TQFT
of \BP\ is. As we will see, it is related to the large $N$ limit
of the $q$-deformed Yang-Mills theory ($q$YM) on $\Sigma$.

By computing the open topological A-model amplitudes on a few
Calabi-Yau manifolds, \BP\ get all others by gluing. The basic
building blocks in \BP\ which we will need are the ``Calabi-Yau
caps'' denoted by $C^{(-1,0)}$ and $C^{(0,-1)}$ carrying
the first Chern classes $(-1,0)$ and $(0,-1)$ respectively
and the pant $P^{(0,1)}$ with $(0,1)$
(see figure 1).

The cap amplitude $C^{(-1,0)}$ is given by
$$
Z^{{\rm top}}(C^{(-1,0)}) =\sum_{R} \; d_q(R)\; q^{-k_R/4}\; \Tr_R U .
$$
Above,
$$
U = P \; e^{i\oint A}
$$
is the holonomy of the gauge field on the D-branes around the circle where D-brane meets $\Sigma$. We take the number of
D-branes to be infinite, so $R$ labels a representation of
$SU(\infty)$. The parameter $q$ is related to the string coupling constant
$g_s$ by
$$
q=e^{-g_s}.
$$
The coefficient $d_q(R)$ is the quantum dimension of the
symmetric group representation corresponding to the Young-Tableaux of
$R$, defined by:
$$
d_q(R) = \prod_{\tableau{1} \in R}{1\over[h(\tableau{1})]_q},
$$
where the product runs over all the boxes in
the Young tableaux of $R$ and $h$ is the hook-length of the
corresponding box. The $q$-analogue $[x]_q$ of $x$ appearing in the formula
is defined by
$$
[x]_q = q^{x/2} - q^{-x/2}.$$
The exponent $k_R$ is given by
$$
k_R =2 \sum_{\tableau{1} \in R} (i(\tableau{1})-j(\tableau{1})),
$$
where $i,j(\tableau{1})$ label the location of the box in the tableaux.
Note that $d_q(R)q^{k(R)/4}$ is the same as the
topological vertex amplitude $C_{R,0,0}$ \AKMV\ with all but one
representation set to be trivial\foot{Our conventions differ
 here from \AKMV\ by $q\rightarrow q^{-1}$.}:
$$d_q(R)q^{k(R)/4}=C_{R,0,0}=W_{R0}.$$
  This is a consistency check, since in this case we are
considering the A-model corresponding to stack of D-branes on Calabi-Yau
$X={\bf C}^3$. This theory is considered in \AKMV\ and shown to
compute $C_{R,0,0}$.

Similarly, using the technology of \BP, we can compute the
Calabi-Yau ``pant'' $P^{(1,0)}$ (see fig. 1). This now carries 3 sets of
representations corresponding to three stacks of D-branes at the 3
semi-infinite tubes with holonomies $U_i$, $i=1,2,3$:
$$
Z^{{\rm top}}(P^{(1,0)}) = \sum_{R} \; {1\over d_q(R)}\; q^{k_R/4}\; \Tr_R
U_1\; \Tr_R U_2\; \Tr_R U_3 .
$$

\bigskip
\centerline{\epsfxsize 3.0truein\epsfbox{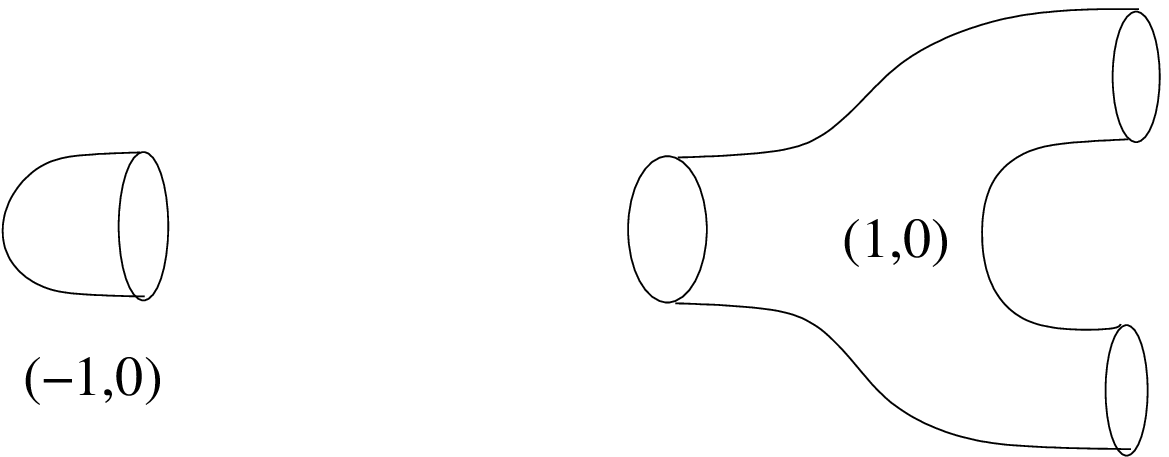}}
\leftskip 2pc
\rightskip 2pc \noindent{\ninepoint\sl
\baselineskip=2pt {\bf Fig.1}
{{The Riemann surfaces corresponding to the cap amplitude
$C^{(-1,0)}$ of the first Chern class $(-1,0)$ and
pant $P^{(1,0)}$ with $(1,0)$.}}}
\bigskip

These, together with the $C^{(0,-1)}$ cap and $P^{(0,1)}$ pant,
$$ \eqalign{&
Z^{{\rm top}}(C^{(0,-1)}) = \sum_{R} \; d_q(R)\; q^{k_R/4} \; \Tr_R U,\cr
&Z^{{\rm top}}(P^{(0,1)}) = \sum_{R}\;{1\over d_q(R)} \;q^{-k_R/4}\;
\Tr_R U_1 \;\Tr_R U_2 \; \Tr_R U_3,}
$$
suffice to compute any Calabi-Yau amplitude with or without D-branes by sewing.

Sewing $\Sigma_L$ and $\Sigma_R$ over
their common boundary we get $\Sigma_{L \cup R}$.
For this, the orientations of the corresponding boundary circles must be opposite. The operation of reversing the orientation of the boundary acts on the holonomy by $U\rightarrow U^{-1}$, and the corresponding path integrals
are sewed together by
\eqn\sewingrule{
Z^{{\rm top}}(\Sigma_{L \cup R})=
\int dU \; Z^{{\rm top}}(\Sigma_L)(U)\; Z^{{\rm top}}(\Sigma_R)(U^{-1}).}
In the representation basis
$$ Z^{{\rm top}}(\Sigma)(U) = \sum_R Z^{{\rm top}}_R(\Sigma)\;
\Tr_R(U),$$
because of the orthogonality of characters
\eqn\orth{
\int dU \; \Tr_{R_1} U \; \Tr_{R_2} U^{-1} = \delta_{R_1 R_2},
}
the sewing in \sewingrule\ corresponds
to identifying the left and the right representation $R$ and summing
over $R$,
$$
Z^{{\rm top}}(\Sigma_{L \cup R})=\sum_{P}\;
Z^{{\rm top}}(\Sigma_L)_P\;Z^{{\rm top}}(\Sigma_R)_P.
$$

Note that, because the pant and the annulus are all diagonal in
representations, the  complex structure moduli of the Riemann
surface one builds by cutting and pasting do not enter in the
resulting topological string amplitudes. This is as it should be,
since these correspond to the
complex structure moduli of Calabi-Yau, and
perturbative A-model topological string amplitudes better not depend on them.

For example, consider the A-model amplitude corresponding to a
Calabi-Yau manifold fibered over a genus $g$ Riemann surface and
bundle with the first Chern class $(2g-2+p, -p)$. A quick counting shows
that we need $(2g-2)$ pants to get a {\it closed} genus $g$ Riemann
surface, by thinking of it as composed of a necklace of $(g-1)$
handle adding operators. We can take
$(2g-2)$ pants of type $P^{(1,0)}$, and to get the bundle right,
insert between them $p$ annuli $A^{(1,-1)}$ obtained by contracting
$C^{(0,-1)}$ and $P^{(1,0)}$. This gives
\eqn\geng{
Z^{{\rm top}}(\Sigma_g) =
\sum_{R} \left( {1\over d_q(R)}\right)^{2g-2} q^{(p+g-1)k_R/2} e^{-t|R|}.
}
Note that in the above formula for $Z^{{\rm top}}$,
$d_q(R)$ is the  $N\rightarrow \infty$ limit of the
quantum dimension ${\rm dim}_q (R)$ of $U(N)$ representation with the same
Young-tableaux. This is suggestive of the relation with
Chern-Simons theory, which we will discuss in section 5.

This answer captures all the non-trivial contributions to perturbative topological
string amplitudes, but does not include contributions from constant maps.  As is well known,
these modify the amplitude by
\eqn\pert{
Z^{{\rm top}}(q,t)\rightarrow Z_0(q,t) \; Z^{{\rm top}}(q,t),}
\eqn\pertii{
Z_0(q,t)= M(q)^{\chi(X)/2}  \;
\exp\left(a\; {t^3\over 6g_s^2} +\,b\;{t\over 24}\right),}
where $M(q)=\prod_{n=1}^{\infty}(1-q^n)^{-n}$ is the McMahon function,
$\chi(X)$ is the Euler characteristic,
$a$ is the triple intersection of the K\"ahler class, and $b$
is related to $c_2$ of the Calabi-Yau $X$.  In the non-compact model
under discussion these are a little ambiguous, but we will find that
the black hole counting agrees with the above if we take
$$\chi(X)=2-2g, \quad a=-{1\over p(p+2g-2)},
\quad  b={p+2g-2\over p},$$
which we will adopt as our definition of perturbative topological string amplitude
in this geometry.  For simplicity of notation we will
drop $Z_0(q,t)$ from the
expressions below, but they should be included when comparing with
the black hole prediction.

As another example, consider again a genus $g$ Riemann surface,
but with $h$ punctures. This corresponds to insertion of $h$
additional pant diagrams. Choosing $(h-r)$ to be of type
$P^{(1,0)}$ and $r$ of type $P^{(0,1)}$ we get a new Calabi-Yau
manifold with fibers of degrees
$$
({\rm deg}(L_1), {\rm deg}(L_2))=(2g-2+h+p',-p'),
$$
where $p'=p-r$
and the topological string amplitude corresponding to it is
$$
Z^{\rm top}(\Sigma_{g,h}) = \sum_{R}\;
\Bigl({1\over d_q(R)}\Bigr)^{2g-2+h}\;q^{{{\rm deg}(L_1)-
{\rm deg}(L_2)\over 4}k_R}\;
e^{-|R| t}\; \Tr_{R}U_1 \cdots \Tr_{R} U_h.
$$

\medskip
\subsec{D-branes in the Fiber}
So far we have considered Lagrangian D-branes wrapping 1-cycles on
$\Sigma$, $i.e.$ D-branes
of topology $S^1\times C$ with the $S^1$ corresponding to a 1-cycle
on $\Sigma$ and $C$ is the 2-dimensional subspace in the fiber.
There is another class of D-branes which will be relevant for us
--  these are Lagrangian D-branes wrapping 1-cycles in the fiber.

\bigskip
\centerline{\epsfxsize 3.0truein\epsfbox{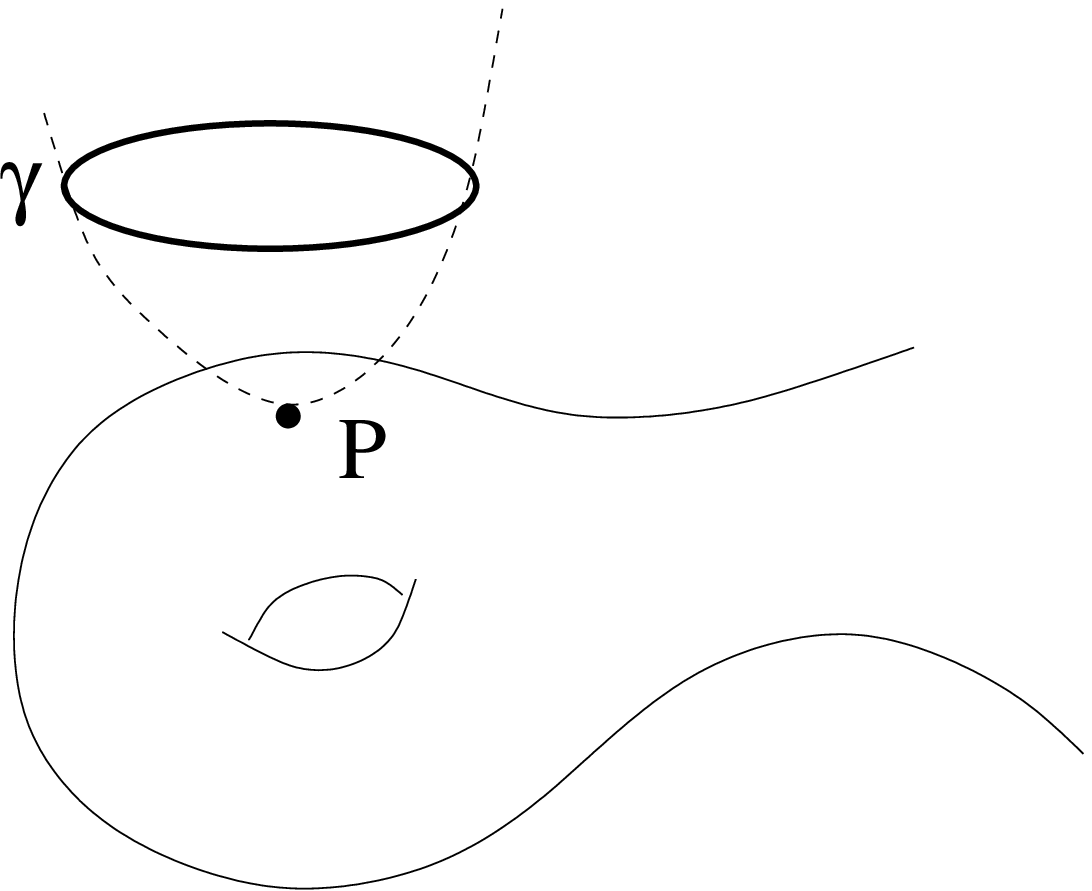}}
\leftskip 2pc
\rightskip 2pc \noindent{\ninepoint\sl
\baselineskip=2pt {\bf Fig.2}
{{The figure depicts a D-bane wrapping Lagrangian cycle $\gamma$ in fiber over a point $P$ on the Riemann surface. Note that the D-brane need not touch the Riemann surface, rather there is a modulus corresponding to
its position in the fiber.}}}
\bigskip
Consider as
before $X= {{L_1\oplus L_2} \rightarrow \Sigma}$ for a fixed Riemann
surface $\Sigma$. Pick a point $P$ on $\Sigma$, and let the D-brane lie
in the fiber above this point. Let $z$ be a local coordinate centered at $P$.
If $(u_1,u_2)$ are coordinates on the fiber $L_1\oplus L_2$ over $P$, then
$(z,u_1,u_2)$ parameterizes a local ${\bf C}^{3}$  patch on $X$.
In this patch there are Lagrangian D-branes of topology $C\times S^1$
where the $S^1$ corresponds to $|u_1|^2=const$, for example. These are
in fact the same D-branes considered in \AKMV , and combining the results of
\BP\ with that of \AKMV , we can also obtain topological string amplitudes
corresponding to them.  The prescription is as follows.

We cut $\Sigma$ into two pieces, a cap
$C_P$ containing $P$ and $\Sigma -C_{P}$. Correspondingly, cut $X$
into two pieces, one corresponding to ${\bf C}^3$ that fibers over
$C_P$, the other to the fibration over $\Sigma-C_P$. We have
$$
Z(\Sigma) = \sum_{R} Z(C_P)_R Z(\Sigma-C_P)_R.
$$
Now, adding a D-brane in fiber over $P$ on $\Sigma$,
will replace $Z(C_P)_R$ by
$$
\sum_Q Z^{top}(C_P)_{RQ}\; \Tr_Q V,
$$ where $V$ is the
holonomy on this D-brane around the fiber $S^1$.  Moreover, this
should have a local effect that can be understood purely in the ${\bf C}^3$
patch we have cut out, and be independent of the rest of $X$.  But,
all the amplitudes with D-branes on ${\bf C}^3$ were computed in \AKMV .
Adding a D-brane in the fiber must correspond, up to framing, to the
topological vertex with one of the partitions trivial, $i.e$,
$$
Z^{top}(C_P)_{RQ}=W_{RQ},
$$
where $W_{RQ}(q)$ is the $N\rightarrow
\infty$ limit of the $S$-matrix of WZW- model and Chern-Simons theory
$$
W_{RQ}(q) = \lim_{N\rightarrow \infty}q^{ N(|R|+|Q|)\over 2}{S_{RQ}(q,N)\over
S_{00}(q,N)}, \quad q=e^{-g_s}.
$$
For example, the amplitude on a genus $g$ Riemann surface with $h$
marked points and
D-branes in the fiber over them is
\eqn\infib{
Z^{top}(\Sigma_{g,h})=\sum_R
{W_{R R_1} \ldots W_{R,R_h}\over W_{R,0}^{2g-2+h}}q^{{{\rm deg}(L_1)-
{\rm deg}(L_2)\over 4}\; k_R}\;
\Tr_{R_1}V_1\cdots \Tr_{R_{h}}V_{h},
}
where we used the fact that $W_{R,0} = d_R q^{k_R\over4}$.
Note that D-branes need not be touching the Riemann surface, but they can
be displaced some distance in the fiber. The modulus corresponding to
where the D-brane is enters replacing
$$
\Tr_R V \rightarrow e^{-s |R|} \; \Tr_R V,
$$
where $s$ is an arbitrary complex parameter. In this case it would
correspond to the area of the disk that attaches the D-brane to the
Riemann surface.

\subsec{An Alternative Derivation}

In this subsection we give a short alternative derivation of the
topological string amplitudes in these backgrounds, analogous to
the derivation of 2d Yang-Mills on a Riemann surface \W . The idea
is that by gluing rules all we need to determine are the cap, the
annulus and the pant diagrams. The cap and the annulus diagram
have been derived before \AKMV (as this is part of the standard
topological vertex construction).  For the pant diagram we would
first argue that ${Z^{top}(P)}_{R_1,R_2,R_3} $ is zero unless all
three representations are the same. As already noted the fact that
the amplitudes of the A-model do not depend on the complex
structure of the Riemann surface implies that
${Z^{top}(P)}_{R_1,R_2,R_3} $ vanishes unless all three
representations have the same number of boxes.  We can also argue
that, more strongly, all three representations have to be the same
for the vertex not to vanish. {}From the quantum foam picture
\ref\IqbalDS{ A.~Iqbal, N.~Nekrasov, A.~Okounkov, and C.~Vafa,
``Quantum foam and topological strings,'' {\tt hep-th/0312022}.
}\ this is obvious because we cannot blow
up the three different legs of the pant differently.  We can
only blow up all of the pant the same way, and this implies
all representations are the same.  Another way to show this,
without using the quantum foam picture, is
to consider a D-brane in the fiber above a point on the pant
near one of the boundaries and study how the amplitude gets modified.
Let us take the fiber D-brane representation to be $Q$.

First let us consider the topological amplitude for the annulus with
representations
$R_1,R_2$ on the annulus and $Q$ in the fiber $Z^{top}(A)_{R_1,R_2,Q}$.  It is easy
to see why this is zero unless $R_1=R_2$.   If this were not
to be the case, there should be a particular `time' along the annulus
(viewed as a cylinder) where the representation changes.  The only
way this can be is at the time corresponding to where the image
of the D-brane is located.  However
 we can move the
image of the fiber D-brane on the annulus without changing the amplitude
because this is just a complex structure deformation.  We thus see that
there is no point where $R_1$ can change to a different representation
$R_2$.  Therefore $R_1=R_2$. Now by putting a cap on one end we obtain
$$Z^{top}(C)_{R} Z^{top}(A)_{R,R,Q}=W_{QR}$$
where the right side is obtained by noting that the new
geometry is exactly what one has in the topological vertex.
Thus we find, using $Z^{top}(C)_{R}=W_{0R}$,
$$Z^{top}(A)_{R_1,R_2,Q}=\delta_{R_1,R_2} {W_{QR_1}\over W_{0R_1}}.$$

Armed with this result let us now consider the amplitude
for the pant with three representations $R_i$, $i=1,2,3$ together with
the fiber representation $Q$,
$Z^{top}(P)_{R_1,R_2,R_3,Q}$. Again we exploit the crucial
point that moving the projection of the fiber D-brane
on the pant is a change
in complex structure and does not affect the A-model
amplitudes.  Moving the fiber brane towards the $i$-th boundary
and using the gluing rule and the above result for the annulus
we see that
$$Z^{top}(P)_{R_1,R_2,R_3,Q}={W_{QR_i}\over W_{0R_i}}Z^{top}(P)_{R_1,R_2,R_3}$$
In other words, putting the fiber D-brane in representation $Q$
 multiplies the amplitude by $W_{QR_i}/W_{0R_i}$
if we are near the $i$-th boundary of the pant.
 However, we can also move the fiber
D-brane to the other boundaries of the pant without changing
the amplitude.  Thus $W_{QR_i}/W_{0R_i}$ should be
the same for all $i$, and arbitrary $Q$ which implies that the $R_i$
are all equal.  Thus all we have
to find then is the representation dependence of the ${Z^{top}(P)}_{R,R,R}$.
But this is easy, because gluing the cap should give
the annulus, which implies
$Z^{top}(C)_{R}{Z^{top}(P)}_{R,R,R}=1$ and leads to
${Z^{top}(C)}_{R,R,R}=1/Z^{top}(C)_{R}=1/W_{0R}$.
Here we have suppressed the framing dependence of the vertex which
can be easily restored.

\newsec{Black Hole and Calabi-Yau $L_1\oplus L_2\rightarrow\Sigma_g$. }

In this section we consider IIA string theory compactified on
Calabi-Yau manifold $X$,
$$
X=L_1 \oplus L_2 \rightarrow \Sigma_g,
$$
of the type studied in the previous section in the context of the
topological string theory. We will take $\Sigma_g$ to be
a genus $g$ Riemann surface, and
$$
{\rm deg}(L_1)=p+2g-2,\quad\quad {\rm deg}(L_2) = -p.
$$
We are interested in counting bound states
of D4, D2 and D0 branes, where the D4-branes wrap
$$
C_4=L_2 \rightarrow \Sigma_g,
$$
and D2 branes wrap $\Sigma_g$.  We fix the number of D4
branes to be $N$ and would like to count the ensemble of bound states
on it.  This can be done by studying the field theory on the brane and
introducing certain interactions on it, which correspond to turning on
chemical potentials for D2 and D0 branes.

The relevant quantum field theory has been studied in \V .  The case
of interest in \V\ was $\Sigma =T^2$ where the ${\cal N}=4$ topologically
twisted gauge theory on $C_4$
was reduced to 2d Yang-Mills gauge theory on $T^2$. The construction in \V\ goes through for any
genus Riemann surface, and we find that the theory reduces
to a gauge theory on $\Sigma$, with one subtlety:  The measure
on field space in this reduction leads to a q-deformed 2d Yang-Mills
theory on $\Sigma$.  This measure does not affect the $T^2$ case studied
in \V\ and affects the partition function only for $g\not=1$.
We will therefore briefly review the construction in \V\ and
take into account the non-trivial measure factor for $g\not=1$.

The world-volume gauge theory on the N D4-branes is the
${\cal N}=4$ topological $U(N)$ YM on $C_4$.
Turning on chemical potentials for $D0$ brane and $D2$ brane
correspond to introducing the observables
\eqn\start{
S_{4d}=\frac{1}{2g_s} \int_{C_4} \tr F \wedge F +
\frac{\theta}{g_s} \int_{C_4} \tr F \wedge K .}
where $K$ is the unit volume form of $\Sigma.$
The parameters $g_s$ and $\theta$ are related
to the chemical potentials $\phi^0$ and $\phi^1$
for $D0$ and $D2$ branes respectively as
\eqn\induced{
\varphi^0=\frac{4\pi^2 }{g_s},\quad \varphi^1=\frac{2\pi \theta }{g_s}.
}

The topological theory is invariant under turning
on a certain massive deformation which simplifies the theory.
By using further
deformation which correspond to a $U(1)$ rotation on the
fiber,\foot{As discussed in \Nekrasov\
such equivariant deformations do not affect ${\cal N}=4$ Yang-Mills
amplitudes, due to the high supersymmetry.}
the theory localizes to modes which are invariant under
the $U(1)$
and effectively reduces the theory to a gauge theory on $\Sigma$.
Let $z$ and $u$ be coordinates on $C_4$ corresponding to
the Riemann surface and the fiber respectively, and let
\eqn\phidef{
\Phi(z)=\int_{S^1_{z,|u|=\infty}} A.}
Here $S^1_{z,|u|=\infty}$ is the circle at infinity in the fiber
over the point $z$ on $\Sigma$ and $A$ is the gauge field
on the world-volume of the D4-branes, so the field $\Phi$ parameterizes the holonomy of the gauge field at infinity.

In reducing \start\ to 2d theory we should take into account two
important effects. First, as shown in \V,
the non-triviality of the fibration $L_2 \rightarrow \Sigma$
generates the following term in the effective 2d action:
\eqn\gener{
\delta S=-\frac{p}{2g_s}\int_{\Sigma}  \tr\Phi^2,}
where $-p$ corresponds to the degree of $L_2$.
The action becomes the action of $U(N)$ 2d YM theory
\eqn\action{
S =\frac{1}{g_s} \int_{\Sigma} \tr \Phi \wedge F +
\frac{\theta}{g_s} \int_{\Sigma} \tr \Phi\wedge K -\frac{p}{2g_s}\int_{\Sigma} \tr\Phi^2.
}

There is however an important subtlety: the field $\Phi$ comes from
the holonomy of the gauge field at
infinity, so it is periodic. More precisely, it is not $\Phi$ but $e^{i\Phi}$
which is a good variable. This does not affect the action, but it
does affect the measure of the path integral. As a consequence, our theory
is a certain deformation of 2d YM, which can be naturally interpreted
as a $q$-deformed version of the Yang-Mills theory as we shall see below.
In the following we will provide two complementary ways to understand this
theory. The first relies on localization and computes the path integral directly.
The second uses an operatorial approach, where we cut $\Sigma$
into pieces on which we can solve the theory.

\subsec{The $q$-Deformed 2d YM Theory I} In this section we use
topological invariance and localization to compute the path integral
directly. We will begin by considering the ordinary 2d YM, and then
show how this gets modified in our case.  We call our theory a
$q$-deformed 2d YM, for the reason which will soon be 
apparent\foot{After completing this paper we became aware 
that q-deformed YM theory was 
studied previously in \roche , \Klim .} .

Underlying the physical 2d YM theory at finite $p$ and $\theta$ is a
topological YM theory \W\  which is in fact the version
of the 2d YM theory arising naturally for us in the reduction
of ${\cal N}=4$ topologically twisted Yang-Mills theory on $C_4$.  To solve
the 2d YM theory it is convenient to use this topological formulation,
as was done in \W\ and \BT\ in two different approaches. We will briefly
review both approaches below, only emphasizing how the change in measure for
$\Phi$ affects the result.  We first follow the approach in \BT\ and in the
following subsection that of \W .

  This
topological YM theory has a BRST multiplet $(A,\psi, \phi)$ with
the path integral
$$
\int {\cal D}A {\cal D}\phi {\cal D}\psi\;
\exp\left[-{1\over g_s}\int_{\Sigma} \tr \left(\phi F + \psi \wedge \psi\right)\right]
$$
and transformation laws
\eqn\BRST{\eqalign{
\delta A &= \epsilon\; \psi ,\cr
\delta \psi &= \epsilon\; D \phi ,\cr
\delta \phi &=  0.
}}
The field $\psi$ is an anti-commuting, Lie-algebra valued
one form on $\Sigma$. Since it is appears quadratically in the action, it is usually neglected.
The transformation laws imply that path-integral localizes on configurations
with $\phi$ covariantly constant $D\phi=0$, where
$\phi$ and $A$ can be simultaneously diagonalized.
We will use this to find an Abelianization of the 2d YM, and
then of our deformation of it. This has been done in \BT ,
and we will sketch the main steps.

Write
$$
\phi=\phi_{i}{T_i} +{\phi}_{\alpha}T_{\alpha},\quad\quad
 A=A_{i}{T_i} +{A}_{\alpha}T_{\alpha},
$$
where $T_i$ are Cartan sub-algebra generators and $T_\alpha$
correspond to roots. We go to the gauge where $\phi$ is diagonal
by setting $\phi_{\alpha}$ to zero. This gives Abelian 2d YM with
an additional coupling to off-diagonal modes of $A$:
$$
\delta S_0=\int_{\Sigma} \sum_{\alpha} \alpha(\phi) A_{\alpha} A_{-\alpha}.
$$
Moreover, to fix the gauge properly, we must also take care of
the path integral measure. This is done by introducing a pair of
ghosts $(c_{\alpha},{\bar c}_{\alpha})$ with coupling
$$
\delta S_1 =\int_{\Sigma} \sum_{\alpha} \alpha(\phi) {\bar c}_{-\alpha}c_{\alpha}.
$$
Integrating over the $c$'s and the $A$'s gives a ratio of one-loop determinants
of a complex scalar of fermionic statistics and commuting one-form which is
hermitian.  The non-zero modes cancel between them \BT ,
and the zero modes give:
$$
{\Delta(\phi)^{2 b_0(\Sigma)}\over \Delta(\phi)^{b_1(\Sigma)}} =
\Delta(\phi)^{\chi(\Sigma)},
$$
where
\eqn\de{
\Delta(\phi) =
\prod_{1\le i < j \le N} (\phi_i-\phi_j).
}
Note that for $g>1$ this appears singular
at places where $\Delta_H(\phi)$ vanishes. When $\Delta(\phi)$
vanishes the gauge group becomes non-Abelian, so
we have integrated out a
massless field, and the singularity is a consequence of it.
In fact \BT\ the singular points give no contribution to the path integral.
By giving a bare mass term to $A_{\alpha}$ we can regularize the path integral
$\sum_{\alpha>0} \mu A_{\alpha}A_{-\alpha}$ it is easy to see that the problematic points give a vanishing contribution for any non-zero value of $\mu$, and by continuity for $\mu$ zero as well.

It is now simple to turn $p$ and $\theta$ back on,
to the physical theory, viewing it as inserting
a $Q$-closed (but not exact) observable to the path
integral of the topological theory.
%
Now, exactly the same discussion would go through
in our case where $\phi$ is compact, however we have to make sure
that the terms we add in $\delta S$ respect this. The easiest way to do this
is by the method of images, more precisely
by adding an infinite sequence of multiplets $(c_{\vec n},\bar c_{\vec n})$,
which couple to ${\vec{\phi} + 2\pi \vec{n}}$.
The effect of this is to replace $\Delta(\phi)$ by
\eqn\deH{
\Delta_H(\phi) =
\prod_{1\le i < j \le N}2 \sin\left({\phi_i-\phi_j\over 2}\right).
}
To summarize, we end up with an Abelian gauge theory:
$$
Z^{qYM}(\Sigma)=
{1 \over N!}\int'  \prod_i
d\phi_i \Bigl[\Delta_H(\phi)\Bigr]^{\chi(\Sigma)}
\exp\left[\int_{\Sigma}\frac{p}{2g_s} \sum_i\phi_i^2 -
\frac{\theta}{g_s} \sum_i \phi_i-\frac{1}{g_s} \sum_i  F_i \phi_i  \right],
$$
where $\int'\prod_i d \phi_i$ denotes the path integral with points with
$\Delta_H(\phi)=0$ omitted. The ${1\over N!}$ factor corresponds to dividing
by the volume of the Weyl group, as a discrete permutation symmetry
inherited from $U(N)$. The normalization of the
path integral has ambiguities coming in part from the choice of regularization.
As explained in \W, for the theory at hand, which is essentially
topological or more precisely invariant under area preserving
diffeomorphisms of $\Sigma$, the different regularizations differ by
additions of terms of the form
\eqn\ambig{
a \int_{\Sigma} R + b \int_{\Sigma} K = a \chi(\Sigma) + b,
}
to the action.
For us, the constants $a$ and $b$ will be fixed by
the black hole physics. In the following, we will not worry
about the normalizations until the end of the subsection.

Note that above coincides, at $p=0=\theta$ with the partition
function of Chern-Simons theory on $S^1 \times \Sigma$ derived in
\BT\ by a different, but related method. At more general values of
$p$, this should correspond to the Chern-Simons theory\foot{This theory
has been recently studied in
\ref\BW{C. Beasley and E. Witten, private communication.} with a different
motivation.} on $S^1$ bundle
over $\Sigma$ with the first Chern class $-p$. It should be clear why
the Chern-Simons theory appears: The non-compact four-cycle $C_4$
wrapped by the D4 branes has this as a boundary. The action
\start\ should be viewed as providing a definition of the
Chern-Simons theory for non-integer values of $k$ in $g_s
=2\pi/{(k+N)}$, taking $g_s$ to be in principle arbitrary. We will discuss the
relation of the D4 brane theory on $X$ to the topological string
theory on $X$ in section 5.

We can in fact give a completely explicit expression for the above
path integral.
Integrating over the gauge fields reduces \BT\ to a sum over nontrivial
$U(1)^{N}$ bundles on $\Sigma$. These are classified by their first
Chern classes,
$$
F_i=2\pi r_i  K,\quad r_i \in {\bf Z}.
$$
Summing over all flux configurations, we  find that integral over $\phi_i$ gets
contribution from
$$
\phi_i= i g_s n_i,\quad n_i \in {\bf Z},
$$
so that partition function
is given by\foot{Note that the sign of theta is not meaningful, the theories at $\theta$ and $-\theta$ are equivalent.}
\eqn\ymred{
Z^{qYM}(\Sigma)=
{ 1\over N!}{\sum_{n_i \in {\bf Z}}}' \left( \prod_{1\le i < j \le N}[n_i-n_j]_q
\right)^{2-2g}
q^{\frac{p}{2}C_2({\vec n})}e^{i\theta C_1({\vec n})},
}where
$$
\eqalign{q&= e^{-g_s},\cr
C_2({\vec n})=\sum_{i=1}^{N} n_i^2,& \quad
C_1({\vec n})=\sum_{i=1}^{N}n_i,}
$$
and $[x]_q$ is the $q$-analogue of $x$
defined as usual by $[x]_q=q^{x\over 2}-q^{-{x\over 2}}$.
We must recall that points where $\Delta_H(\phi)=0$ are omitted
from the path integral, and correspondingly we omit terms in the sum
where $n_i=n_j.$ We used $\sum_{n_i \in Z}'$ to denote this modified sum.

In fact, we can rewrite the above as a sum over representations $\R$ of
$U(N)$. By using the Weyl invariance to restrict the fundamental
Weyl chamber, $n_{1}> n_2 > \cdots > n_{N}\geq 0$, and then letting $n_i=\R_i$
where $\R_i$'s label the lengths of rows of $\R$. Finally it is
convenient to shift the $\R_i$ by $\rho_i ={1\over 2}(N - 2i+1)$, $i.e.$ we
shift by the Weyl vector.\foot{More precisely, this is valid only
when $N$ is odd, otherwise
we shift $n_j=\R_j +{N \over 2} -j$ and also introduce a shift of $\theta$
angle.} Then, $C_{1,2}$ become the
corresponding Casimir's of $U(N)$
$$
\eqalign{& C_2(\R )=\k_{\R}+ N \vert \R \vert, \quad
\k_{\R}:=\sum_{i=1}^{N} \R_i(\R_i-2i+1),\cr
&
C_1(\R)=\vert \R \vert,\quad \quad\quad \vert \R \vert:=\sum_{i=1}^N \R_i,}
$$
and the partition sum is expressed as
$$
Z^{qYM}(\Sigma)=\sum_{\R} S_{0 \R}^{2-2g}
q^{\frac{p}{2}C_2(\R)}e^{i\theta C_1(\R),}
$$
where $ S_{0 \R}=S_{00} \ {\rm dim}_q (\R)$
and the quantum
dimension of representation $\R$ is given by
\eqn\dimq{ {\rm dim}_q
(\R)=\prod_{1\le i < j \le N}\frac{[\R_i-\R_j+j-i]_q}{[j-i]_q}
}
(note that $S_{00}$ is the denominator in the above expression).
The reader should note here that we have deliberately distinguished the
representations $R$ of $SU(N)$ from representations
${\cal R}$ of $U(N)$ here. This somewhat technical point
will be of crucial importance in section 5 where we consider the large $N$
limit of the $q$-deformed YM theory.

One can recognize in the above the building blocks of the $U(N)$
Chern-Simons theory:
$$
{\rm dim}_q (\R) = S_{\R0}/S_{00},
$$
where $S_{\R{\cal Q}}$ correspond to entries of the S-matrix of
$U(N)_k$ $WZW$ model, albeit at non-integer value of level $k$.
In fact $k$ here is a pure imaginary number.   In particular
we do not have any truncation of the representations, which is usually
associated to integer $k$.

We finish this subsection by giving the normalized $q$-deformed YM
partition function.
As we mentioned above, the normalization is ambiguous due
terms of the form \ambig, but fixed for us by black hole physics
to be discussed in section 5 to be\foot{See
(5.9)-(5.14) in section 5.}
\eqn\ymrediv{
Z^{qYM}(\Sigma)=\a(g_s,\theta)
\sum_{\R} S_{0 \R}^{2-2g}
q^{\frac{p}{2}C_2(\R)}e^{i\theta C_1(\R)}, }
where
\eqn\alpii{
\a(g_s,\theta)=
q^{ \rho^2 (p+2g-2) \over 2p }
e^{\frac{N \theta^2}{2pg_s}} \  q^{(2-2g)\left(
\rho^2+{N \over 24}\right) } .}


\subsec{The q-Deformed 2d YM Theory II}

We now give an operatorial approach for computing the partition function
whose advantage is that it will make a similarity to the topological
theory of section 2 apparent. We will follow closely the
approach of \W : we first solve the theory on a sphere with three
holes, and then get the rest by sewing.

To begin with note that, just as in  the case of ordinary
2d YM, $\Phi$ and $A$ are canonically conjugate variables.  In
quantizing the theory on a cylinder, we can take as basis of the
Hilbert space to be gauge invariant functions of $A$. These are
given by $\chi_{\R}(U) = \Tr_{\R} U$ where
$$
U = P {\rm exp}\ \, i \oint A.
$$
On these
$\Phi(x)$ acts as $\Phi(x) =  g_s {\del\over \del A(x)}$ or
$$
\Phi_a(x)\; \chi_{\R}(U)= i\,  \chi_{\R}(T_a U),
$$
where $T_a$ are generators of the lie algebra $\Phi(x) = \sum_a \Phi_a(x) T_a$.
Recall that, since we are studying a $U(N)$ gauge theory,
$U$ is unitary and $\R$ labels representation of $U(N)$.

We will first solve the theory in the topological
limit, obtained by setting $p=0$ and then study the consequences
of turning on $p$. For simplicity, we will also turn off the theta
angle and restore it back later.

Consider the path integral on a pant $P$ i.e. a
sphere with three punctures with holonomies $U_i$, $i=1,2,3$ around them. This is of the form
$$
Z(P)(U_1,U_2,U_3) = \sum_{{\R}_1,{\R}_2,{\R}_3}\; Z(P)_{{\R}_1,{\R}_2,{\R}_3}\;
\Tr_{{\R}_1} U_1 \; \Tr_{\R_2} U_2 \; \Tr_{\R_3} U_3.
$$
We will now follow \W\ to argue that $Z(P)_{\R_1,\R_2,\R_3}$ vanishes unless
$\R_1=\R_2=\R_3=\R$.  Inserting an arbitrary operator ${\cal O}(\Phi(x))$
at any one of the punctures picks out the corresponding Casimir of the
representation at that puncture. Using the invariance of the observable on
the point of insertion, we can move the operator ${\cal O}(\Phi)$ to any
other puncture.

By similar argument,
the path integral on an annulus $A$ of length $T$ is given by
$$
Z(A)(U_1,U_2) =\sum_{\R} \;\Tr_\R U_1 \;\Tr_\R U_2.
$$
As annulus amplitude is the propagator, we would usually have
representation $\R$ weighted by $e^{-H(\R) T}$, where $H$ is the
Hamiltonian and $T$ is the length of the propagator. However $H$
vanishes in the topological theory: at $p=0=\theta$.

Finally, consider the path integral on a cap $C$
$$
Z(C)(U) = \sum_\R \;Z(C)_\R\; \Tr_\R U.
$$
First, note that the cap and the pant amplitudes
are not independent. Gluing the cap $C$ to the pant $P$ we must
recover the annulus amplitude:
$$
\int dU Z(C)(U^{-1})\; Z(P)(U,U_1,U_2) =Z(A)(U_1,U_2).
$$
This implies
$$
Z(C)_\R = {1\over Z(P)_{\R}},
$$
where we used a $U(N)$ analogue of the $SU(N)$ orthogonality
relation \orth\
$$
\int dU \;\Tr_{\R} U \;\Tr_{\cal Q} U^{-1} = \delta_{\R,{\cal Q}}.
$$
Now we come to the main point of the discussion,
which is the computation of the cap amplitude of the deformed 2d
YM.

To begin with, recall what happens in ordinary the 2d YM.
Computing a path integral with fixed holonomy $U$ on the boundary
simply gives a delta function on the holonomy $\delta(U-1) =
\sum_{\R} {\rm dim} (\R) \, \Tr_{\R}(U)$, so that
$$
Z^{{\rm 2dYM}}(U)(C) =  \sum_\R {\rm dim} (\R) \; \Tr_\R\,U.
$$

Now, we could have also computed this in the basis where we fix $\Phi(x)$
on the boundary and not $U$. Since $\Phi(x)$ and $A$ are canonically conjugate,
this is related to the above by a simple Fourier transform,
$$
Z^{{\rm 2dYM}}(C)(\Phi)
= \int DU \; e^{{1\over g_s}\oint_{\partial C} \Tr {\Phi A}} \;
Z^{{\rm 2dYM}}(C)(U),
$$
where the path integral is taken on the boundary of the disk.
We will show in the appendix A that the integral is simply given
by
\eqn\jacob{
Z^{{\rm 2dYM}}(C)(\phi)
= \prod_{1\le i < j \le N}{(\phi_j-\phi_i)} =\Delta(\phi),
}
%
%
%
where $\phi_{i}$ ($i=1,\ldots N$) are eigenvalues of $\Phi$.
Note that $\Delta(\phi)$ is the same function \de\ we met in
the previous subsection.

Now, let us return to the deformed YM theory. In this case
$\phi_i$'s are periodic, so the wave function \jacob\ is not well defined.
One can make it well defined by adding images under
$\phi_i \rightarrow \phi_i + 2\pi $.
This, as in the previous subsection, replaces
$\Delta(\phi)=\prod_{1\le i < j \le N}{(\phi_j-\phi_i)}$
with
$$
\Delta_H(\phi)=\prod_{1\le i < j \le N}
2 \sin\left({\phi_i-\phi_j\over{2}}\right).
$$
Alternatively, we can express the wave function in terms of the
holonomy basis, by undoing the Fourier transform. This gives:
$$
Z^{qYM}(C)(U) =  \sum_\R {\rm dim}_q (\R) \; \Tr_\R U.
$$
where the quantum dimension of representation $\R$ is given by \dimq .
Note that ${\rm dim}_q (\R)$ becomes ${\rm dim}(\R)$
in the limit $g_s \to 0.$

Finally, to get back to the physical theory by turning on $p$
and $\theta$. As explained in \W , the difference between the
physical and the topological theory is only that the Hamiltonian of the former
is not vanishing, and it is given by\foot{See footnote 4, on page 15.}
$$
H  = {1\over 2}\,g_s p\; C_2 -i\theta C_1.
$$
This gives the area dependence to the amplitudes. For example,
the amplitude for an annulus of length $T=1$ is
$$
Z^{qYM}(A)(U_1,U_2) =\sum_{\R} \;q^{pC_2(\R)/2} \;e^{i \theta \R} \;
\Tr_\R U_1 \;\Tr_\R U_2.
$$
Since we can change the area of the pant $P$ and the cap $C$ by adding annuli,
the corresponding amplitudes of the physical theory are:
$$
Z^{qYM}(P)(U_1,U_2,U_3) = f \sum_{\R} {1\over {\rm dim}_q(\R)}\;q^{pC_2(\R)/2}\;
e^{i \theta \R}\;\Tr_{\R} U_1 \; \Tr_{\R} U_2 \; \Tr_{\R} U_3,
$$
and
$$
Z^{qYM}(C)(U) = {1\over f}\sum_\R \;{\rm dim}_q(\R)\;q^{pC_2(\R)/2} \;e^{i \theta \R}
\;\Tr_\R U,
$$
where we have normalized
the area of the resulting surface to one.  The normalization
factor $f$, which is independent
of the $U_i$'s, cannot be fixed by this argument and depends on the
normalization of the path-integral.  Consistency with the previous
discussion leads to $f=1/S_{00}$.
{}From these we can get arbitrary amplitudes of our deformed 2d YM
theory on any Riemann surface by sewing.

For example, we can finally compute the
partition function on a genus $g$ Riemann surface. We do so by sewing
$(2g-2)$ pant amplitudes and find
$$
Z^{qYM}(\Sigma_g) =S_{00}^{2-2g}\sum_\R  \left({1\over {\rm dim}_q(\R)}\right)^{2g-2}\;
q^{pC_2(\R)/2}\; e^{i \theta \R},
$$
where we have set the area of the surface to $1$ (we also
need the prefactor $\alpha(g_s,\theta)$, discussed before, in comparing with black-hole
physics).
Note that this amplitude, at $p=0=\theta$, is identical
to that of the Chern-Simons theory on $\Sigma_g\times S^1$.  For general
$p$ but $\theta =0$ this is equivalent to the result for the Chern-Simons theory
on a circle bundle over $\Sigma$ where the circle bundle is identified with
the circles of fixed norm of the $L_2$ bundle over $\Sigma$. The only difference
being that here $k$ is not an integer and we do not have the usual truncation
of representations which occurs for integer $k$.

Finally, in the above, we have mainly been considering wave-functions
in the polarization where we specify $A$ on the boundary. We can rewrite all the
amplitudes corresponding to manifolds with boundaries in terms of keeping
$\Phi=-i g_s {\del\over \del A}$ fixed on the boundary. For example, consider the
pant amplitude.
$$
Z^{qYM}(P)(V_1,V_2,V_3) = \sum_{\R_1,\R_2,\R_3} {Z^{V}}(P)_{\R_1,\R_2,\R_3} \;
\Tr_{\R_1}V_1\;\Tr_{\R_2}V_2\;\Tr_{\R_3}V_3 ,
$$
where, since only $e^{i\Phi}$ is well defined,
$$
V_i =e^{i\Phi_i}
~~~(i=1,2,3).$$
Now, viewed from the three-dimensional perspective,
 we have a three-manifold with three $T^2$ torus boundaries.
Fixing $V=e^{i\Phi}$ on the boundary, fixes
the holonomy around $a$-cycle of the $T^2$ say (corresponding to the $S^1$
fiber over the pant $P$), whereas fixing
$U=Pe^{i \oint A}$ fixes holonomy around the $b$ cycle. Now
we've derived the amplitude corresponding to fixing $U$.
We could compute from it the amplitude with $V$ fixed if we could
exchange the $a$ and the $b$ cycle.
Fortunately, in the Chern-Simon theory, such operation is well known and  is implemented
by the $S$ matrix
$$
S_{\R_1\R_2}
$$
of the $WZW$ model. So we have:
\eqn\dualpant{
Z^{V}_{\R_1,\R_2,\R_3} = \sum_{\R} {S_{\R_1\R}S_{\R_2\R}S_{\R_3\R}\over S_{0\R}}\;q^{pC_2(\R)/2}\;e^{i \theta C_1(\R)}.
}
The right hand side, at $p=0=\theta$
is simply the Verlinde formula, for non-integer $k$.
This of course is not a surprise: in that case, we
are effectively computing a Chern-Simons amplitude on $S^1\times S^2$
with Wilson-lines on $S^1$ in representation $\R_1,\R_2,\R_3$ over three
points in the $S^2$, and the results are the fusion coefficients,
analytically continued to arbitrary $k$ using the Verlinde formula
\JP.

\newsec{Black Hole Entropy}

In this section we will count the number of BPS states of
the black hole that arises
from $N$ D4-branes on $C_4$ with any number of D0-branes and
D2-branes wrapping $\Sigma$.  As we will note below, in order
to do this we have to perform a modular transformation on the partition function
evaluated in the previous section.

In the topologically twisted ${\cal N}=4$ YM theory discussed
in the last section we turned on observable corresponding to D-brane charges
$$
S_{4d}=\frac{1}{2g_s} \int_{C_4} \tr F \wedge F +
\frac{\theta}{g_s} \int_{C_4} \tr F \wedge K ,
$$
where $K$ is K\"ahler class of $\Sigma_g$.
The $D_0$ and $D_2$ brane charges $q_0, q_1$ are
measured by
\eqn\topclasses{q_0 = \frac{1}{8\pi^2}\int_{C_4} \tr F\wedge F, ~~
q_1=\frac{1}{2\pi} \int_{C_4}  \tr F \wedge K.}
It was shown in \VW\ that the functional integral
$$
Z^{q{\rm YM}} = \int {\cal D} A \exp\left(-\frac{1}{2g_s} \int_{C_4} \tr F \wedge F -
\frac{\theta}{g_s} \int_{C_4} \tr F \wedge K \right)
$$
with an appropriate gauge fixing has an expansion of the form,
\eqn\vafawitten{
Z^{q{\rm YM}} = \sum_{q_0, q_1} \Omega(q_0, q_1; N) \exp\left[
-{4\pi^2 \over g_s} q_0 - {2\pi \theta \over g_s} q_1 \right],}
where $\Omega(q_0,q_1; N)$ is the Euler characteristic of the
moduli space of $U(N)$ instantons in the topological
sector set by \topclasses\ and it can also be regarded as the
Witten index for the black hole with the given D-brane charges.
By setting
$$ \varphi^0 = {4\pi^2 \over g_s}, ~~~ \varphi^1 = {2\pi \theta\over g_s}, $$
the instanton expansion of $Z^{q{\rm YM}}$ can be expressed as
\eqn\instantonexp{
Z^{q{\rm YM}} = \sum_{q_0,q_1} \Omega(q_0,q_1;N)
\exp\left[ -q_0\varphi^0-q_1\varphi^1\right].}
For the non-compact manifold $C_4$ ,
the charges $q_0, q_1$ can
be fractional. We will find below that $q_0 \in {1\over 2p}{\bf Z}$
and $q_1 \in {1\over p} {\bf Z}$.

The expressions we gave for $Z^{q{\rm YM}}$ in the previous section,
however, are expansions in $q=e^{-g_s}$ as in \ymred\foot{We have chosen
the normalization of the $q{\rm YM}$ path integral to include the prefactor
$\a(g_s,\theta) \ q^{-{p\rho^2 \over 2}}.$  This
choice is  required for the factorization into chiral blocks in section 5.}
\eqn\dem{\eqalign{
Z^{q{\rm YM}}(\Sigma_g)=&  \a(g_s,\theta) \ q^{-{p\rho^2 \over 2}} \cr
& \times {1 \over N!}\sum'_{ {\vec n} \in {\bf Z}^N}
\left( \prod_{i<j}[n_j-n_i]_q\right)^{2-2g} \;
q^{{p\over 2}{\vec n}\cdot {\vec n}} \;e^{i\theta {\vec n} \cdot {\vec e}},
}}
where $\sum'_{{\vec n}}$ the modified sum omitting $n_i = n_j$ for $i \neq j$,
$\a(g_s,\theta)$ is defined in \alpii, and ${\vec e}=(1,1,\ldots,1).$
Fortunately, the beautiful fact that ${\cal N}=4$
YM theory has $S$-duality saves the
day. The $S$-duality implies that $Z^{q{\rm YM}}(g_s,N)$ is a modular form, which
turns the expression \dem\ into the form \instantonexp.
We will now demonstrate this.

Let us first review the genus $g=1$ case studied in \V .
For $g=1$, the partition sum can be expressed in terms of
elliptic functions \Rudd\ as in
\eqn\above{
Z^{q{\rm YM}}(T^2)={q^{{N\theta^2\over 2 p g_s}
+{(1-p)\over 2}\rho^2}\over N!}
\left[ \vartheta(z,\tau)^N -\frac{N!}{2!(N-2)!}\vartheta(z,\tau)^{N-2}
\vartheta(2z,2\tau)-\cdots \right],
}
where
$\vartheta(z,\tau)=\sum_{n\in Z} e^{\pi i \tau n^2 +2i\pi z n}$ and
$$
z = {\theta\over 2\pi},\quad\quad \tau = {i \over 2\pi} pg_s .
$$
Note that the first term $\vartheta(\z,\tau)^N$ in the right-hand side
of \above\ is obtained from \dem\ by ignoring the constraints $n_i \neq n_j$
for $i\neq j$, and the other terms $\vartheta(z,\tau)^{N-2}\vartheta(2z,2\tau)
+\cdots$
are corrections to incorporate these
constraints. Applying the
modular transformation,
\eqn\thetamodular{\vartheta(z,\tau)=
{1\over \sqrt{-i \tau}}e^{-i\pi {z^2\over \tau}}
\vartheta\left({z\over \tau},-{1\over \tau}\right),}
to the first term \above, we find
\eqn\transfabove{
Z^{q{\rm YM}}(T^2)= {q^{{(1-p)\over 2}\rho^2}  \over N!} \left(
{2 \pi \over p g_s}\right)^{N\over 2}
\sum_{\vec m \in {\bf Z}}
e^{-{1\over 2p}{\vec m}^2\varphi^0+
{1\over p}{\vec m} \cdot {\vec e}\varphi^1}+\cdots.
}
The other terms can be transformed similarly.
Comparing this with \transfabove , we find that $q_0,q_2$ are fractions
$q_0 \in {1\over 2p} {\bf Z}, q_1 \in {1\over p}{\bf Z}$
and that the Witten index of black hole states is given by
$$\Omega(q_0,q_1;N) =\# \left| \{~ {\vec m} \in {\bf Z}^N ~|~
q_0 = {1\over 2p} {\vec m}^2 , ~ q_1= {1\over p}{\vec e}\cdot {\vec m} \}
\right|
+ \cdots.$$

For $g=0$, things are simpler since the factor $\prod_{i<j} [n_i - n_j]_q^2$
automatically takes into account the constraints $n_i \neq n_j$ ($i\neq j$) and
we may extend summation over unrestricted $n_i \in Z$ without any subtractions.
It is useful to expand this factor
as a sum over the Weyl group $W$ of $U(N)$ as in
\eqn\szero{
\prod_{1\le i < j \le N}[n_i-n_j]_{q}=
\sum_{w \in W} \e(w)  q^{- w(\rho)\cdot  n },
}
where $\e(w)=\pm 1$ is the parity of the Weyl group
element $w$. We can then express the partition sum as
\eqn\newym{
Z^{q{\rm YM}}(S^2)=
{q^{{N\theta^2\over 2 p g_s}+{N\over 12}-{(p^2-5p+2)\over 2p} \rho^2}
\over N!}\sum_{w,w' \in W}
\e(w)\e(w') \prod_{k=1}^N \vartheta(z_k(w,w'),\tau)}
where
\eqn\deni{
\tau={i\over 2 \pi}p g_s ,\quad z_k(w,w')= {1\over 2\pi}\left(
\theta  - i g_s a_k(w,w')\right) ,}
and
$$a_k(w,w')=
w(\rho)_k+w'(\rho)_k.$$
Now we use the modular transformation \thetamodular\ of $\vartheta(z,\tau)$
to recast $Z^{q{\rm YM}}(S^2)$ in the form:
\eqn\recast{
\eqalign{ Z^{q{\rm YM}}(S^2)=&
 {q^{{N\over 12}-{(p^2-5p+2)\over 2p} \rho^2}\over N!}\
\left( {2\pi \over pg_s}\right)^{N/2}\sum_{w,w' \in W}\e(w)\e(w')\cr
&\times \sum_{{\vec n} \in {\bf Z}^N}
\exp\left[ -\frac{2 \pi^2}{pg_s}\left( {\vec n} +
{i g_s\over 2\pi} {\vec a}(w,w')\right)^2 +
 {2\pi \theta\over pg_s}{\vec n}\cdot {\vec e}  \right].}
}

It is instructive to compare this with mathematical results
on the Euler characteristic
 of the moduli space of instantons on the four-manifold
$C_4 = {\cal O}(-p) \rightarrow {\bf P}^1$.

When $p=1$, the partition function \recast\ can be
expressed as
\eqn\finform{ Z^{q{\rm YM}} = f(\varphi^0) \left[ \sum_n e^{
-{1 \over 2}n^2 \varphi^0 - n \varphi^1} \right]^N,}
for some $\varphi^0$ dependent factor $f(\varphi^0)$.  The prefactor
would be related to the bound state of the D0 brane to the D4 brane
which is ambiguous in the present context due to the non-compactness
of the D4 brane. (In the compact case it would have been
$\eta^{-N \chi(C_4)}(\varphi_0)$.)
However the $D2$ branes bound to D0 and D4 branes are unambiguous because
they are frozen on the compact part of the geometry, which is the Riemann
surface. In this case, $C_4$ is the total space of ${\cal O}(-1)$ over
${\bf P}^1$, which is a blowup of ${\bf C}^2$ at one point, and \finform\
exactly reproduces the $\varphi^1$ dependence of the blow-up formula
conjectured in \VW\  and proven in \yoshioka.

When $p=2$, \recast\ becomes
\eqn\pistwo{ \eqalign{
Z^{q{\rm YM}}
&= {q^{\rho^2+ N\over 12}\over 2N!}
\left( {2\pi \over pg_s}\right)^{N/2} \sum_{w,w'\in W} \epsilon(w)\epsilon(w')
   q^{-{1\over 2p}a(w,w')^2}
\cr &~~~~~~~~~~~~~~\times
\sum_{{\vec n} \in {\bf Z}^N}(-1)^{{\vec n} \cdot {\vec a}(w,w')}
e^{ -{1\over 4} {\vec n}^2 \varphi^0 - {1\over 2} {\vec e}
\cdot {\vec n} \varphi^1}\cr
&= {q^{\rho^2+ N\over 12}\over 2N!}
\left( {2\pi \over pg_s}\right)^{N/2}
\sum_{w,w' \in W} \epsilon(w) \epsilon(w') q^{{1\over 4\pi p}a(w,w')^2}
\cr &~~~~~~~~~~~~~~\times
\prod_{i=1}^N
\left[ \sum_{r=0,{1\over 2}} (-1)^{2r a_i(w,w')}
\sum_{n\in {\bf Z}}
e^{ - (n+r)^2 \varphi^0 -
(n+r) \varphi^1 }\right].}}
Note that the characters of the level $1$ $SU(2)$ affine Lie algebra are
given by
$$ \chi_r^{{\rm level}~ 1}(\varphi^0,\varphi^1) = {\sum_n
e^{ - (n+r)^2 \varphi^0 -
(n+r) \varphi^1 } \over \prod_{n=1}^\infty
(1-e^{-n\varphi^0}) },$$
where $r=0,{1\over 2}$ corresponds to the spin $0$ and ${1\over 2}$
 representations
respectively. Since \pistwo\ shows that
the $\varphi^1$-dependence of $Z^{q{\rm YM}}$ is
given by a product of $N$ of such characters, one can expanded
$Z^{q{\rm YM}}$ as a sum over the characters of the $SU(2)$ affine
Lie algebra of level $N$
with $\varphi^0$ dependent coefficients.
This agrees with the result by Nakajima \nakajima\ that  the
level $N$ affine algebra acts on the cohomologies
of the moduli space of $U(N)$ instantons on ${\cal O}(-2)\rightarrow {\bf P}^1$,
which is our $C_4$ in the case of $p=2$. The choice of
the $SU(2)$ representation is determined by the boundary condition at
the infinity. Since the $S$-duality transformation mixes up
the boundary conditions, it is reasonable that $Z^{q{\rm YM}}$ computed
in the previous section becomes a sum of the affine $SU(2)$ characters
after the $S$-dual transformation.

Thus, we have demonstrated for $g=0$ and $1$
that our computation of $Z^{q{\rm YM}}$ based on the
reduction to the $q$-deformed $2d$ YM on $\Sigma_g$ agrees
with the instanton expansion of the ${\cal N}=4$ YM on $C_4$
and that $Z^{q{\rm YM}}$ is indeed the generating function of the BPS
black hole that arises from wrapping D4-branes on $C_4$.
It would be interesting to test this for $g \geq 2$ also.
Our next task, however, is to relate $Z^{q{\rm YM}}$ to {\it closed}
topological string
amplitudes on the Calabi-Yau manifold $X$.

\newsec{Large $N$ Limit of $Z^{q{\rm YM}}$ and the Relation to Perturbative Topological
Strings}

The deep relation, conjectured in \osv ,
between 4-dimensional black holes and topological strings
predicts that the partition function of black holes on $X$ $Z_{BH} = Z^{q{\rm YM}}$,
for large charges, is related to the perturbative topological string partition function $Z^{{\rm top}}$ on $X$ as
\eqn\topstringfactorize{
Z^{q{\rm YM}} \sim Z^{{\rm top}}{\bar Z}^{{\rm top}}.}
In this section we will aim to get a better understanding of
this relation by considering the large $N$ ($i.e.$ large black hole charge)
limit of the quantum $2d$ YM
theory on a Riemann surface $\Sigma$ of genus $g$.

The conjecture of \osv\ predicts a precise relation of the parameters of the
YM theory describing the black hole and the dual gravity, or topological
string theory.
According to the conjecture \osv , when one relates the black hole
to the topological string theory, the moduli of the
Calabi-Yau manifold are fixed by the black hole attractor mechanism.
This fixes the real parts of the
projective coordinates $(X^0, X^1)$ on Calabi-Yau moduli space
to magnetic charges of
D6 and D4 branes and their imaginary parts are the chemical
potentials $\varphi^0, \varphi^1$ for the electric D0 and D2 brane charges.
In the current setup, there is no D6 brane magnetic charge, and we have $N$ D4 branes.  Now, the magnetic charge for a single D4 brane, if we measure
in terms of electric units of D2 branes wrapping $\Sigma$, is given by
the intersection number of $\Sigma$ and the 4-cycle the D4 brane wraps.
In the present case this is
$$\#(\Sigma \cap C_4)=p+2g-2,$$
as can easily be seen by deforming $\Sigma$ away from $C_{4}=L_2
\rightarrow \Sigma$ using a
generic section of $L_1$.
Thus, in the black hole background, the projective moduli for the {\it closed} topological
string are fixed to be
\eqn\attractor{
  X^0 = i{\varphi^0 \over \pi},~~
X^1 = (p+2g-2) N - i{\varphi^1 \over \pi}.}
In the previous section, we saw that the chemical potentials are related to the
$q$YM parameters as
$ \varphi^0 = {4\pi^2 \over g_s},$ and $\varphi^1 = {2\pi \theta \over g_s},$
so that
$$ X^0 = {4\pi i \over g_s}, ~~
X^1 = (p+2g-2) N - 2i{\theta \over g_s}.$$
Since the K\"ahler modulus $t$ corresponding to the
base $\Sigma$ of the Calabi-Yau manifold
is given by
$$t = 2\pi i {X^1 \over X^0}$$
in terms of projective coordinates, we
expect that the topological closed string theory
which is the gravity dual of the $U(N)$ 2d
$q$YM theory has $t$ fixed to be
\eqn\kahlermoduli{
t=\half (p+2g-2)Ng_s -i\theta.}

The large $N$ limit of the ordinary 2d YM theory, a cousin of our theory,\foot{
Since $q=e^{-g_s}$ and $g_s \sim g_{{\rm YM}}^2$, the ordinary 2d YM theory
with finite $g_{{\rm YM}}$ is not a limit of our $q$-deformed theory.}
was studied in \gt.
They find that, at large $N$, the Hilbert space of the YM theory factorizes as
$$
{\cal H}_{{\rm YM}}
\rightarrow {\cal H}_{chiral}\otimes{\cal H}_{antichiral}
$$
with ${\cal H}_{chiral}$ corresponding to representations
$\R_+$ with much less than $N$ boxes, and ${\cal H}_{antichiral}$
corresponding to
representations $\R_{-}$ of order $N$ boxes. The same will apply
to the $q$-deformed YM theory.
Corresponding to this, the partition function $Z^{q{\rm YM}}$ should
factorize as
\eqn\qymfactorize{
Z^{q{\rm YM}} \sim \;Z^+_{q{\rm YM}} \;Z^{-}_{q{\rm YM}}.}
It is natural to expect that the two factorizations \topstringfactorize\
and \qymfactorize\ are related. More precisely, one would expect that
chiral $q$YM
partition function $Z^{+}_{q{\rm YM}}$ can be written as a holomorphic
function of $t$, and identified with
the topological string amplitude $Z^{{\rm top}}(t)$
$$
Z^{+}_{q{\rm YM}}(N,\theta,g_s) \sim Z^{{\rm top}}(t,g_s).
$$
We will show that this is indeed the case in the class of Calabi-Yau
manifolds at hand, with some important subtlety that we will describe
in detail below.

The cases $X=L_1\oplus L_2
\rightarrow \Sigma_g$ for $g\geq 1$ and
$g=0$ work somewhat differently in their technical aspects, so we will
consider them separately below.

\subsec{Factorization into Chiral Blocks for Genus $g>1$ Case.}

We consider the large $N$ limit of the BH partition
function \ymrediv\ for $g>1$
$$
Z^{q{\rm YM}}(\Sigma_g)=\a(g_s,\theta) \sum_{\R} S_{0\R}^{2-2g}
q^{\frac{p}{2}C_2(\R)}e^{i\theta C_1(\R)}.
$$
As we will show below in this section the natural value of the
normalization constant $\a(g_s,\theta)$ is as defined in \alpii.  Here
$\R$ labels $U(N)$ representations. It will be more convenient use
the decomposition of $\R$  in terms of $SU(N)$ representation
$R$ and the $U(1)$ charge $m$. First, we will have to recall
how various quantities pertaining to $U(N)$ relate to those of
$SU(N)$ and the $U(1)$.

The Young-tableaux of $\R$ differs from that of $R$ by having
$r$ columns of length $N$ attached: \eqn\decomp{\eqalign{
&\R_i=R_i+r,\quad i=1,\ldots, N-1 \cr &\R_N=r.}} The $U(1)$ charge
is then given by
$$
m=\vert R \vert + Nr, \quad r \in {\bf Z}.
$$

The $U(N)$ Casimir's
are related to the $SU(N)$ Casimir's as
\eqn\kas{
C_1(\R)=m,\quad C_2(\R)=C_2(R)+\frac{m^2}{N},
}
where
\eqn\sun{
C_2(R)=\k_R + N\vert R \vert-\frac{\vert R \vert^2}{N},\quad
\k_R=\sum_{i=1}^{N-1} R_i(R_i-2i+1)
}
Their  quantum dimensions are equal
$${\rm dim}_q (\R) ={\rm dim}_q (R)=\prod_{1
\le i<j \le N} {[R_i-R_j+j-i]_q\over [j-i]_q }.$$ Just as for
ordinary $2d$ YM \gt, the factorization of the Hilbert space at
large $N$ is captured by writing an irreducible $SU(N)$
representation $R$ in terms of the coupled representations
$R_{+}\bar R_{-}$ with $R_{+}$ and $R_{-}$ labelling states in
${\cal H}_{chiral}$ and ${\cal H}_{antichiral}$, respectively. The
Casimir's decompose as follows: the $U(1)$ charge $m$ of $R$
becomes
$$
m=Nl+\vert R_{+}\vert - \vert R_{-} \vert,
$$
where $l=r+R_{-,1}$, and
\eqn\seckas{
C_2(R_{+}\bar R_{-})= C_2(R_+)+C_2(R_-) + 2 {\vert R_+\vert \vert R_-\vert\over N}.}
%
\bigskip
\centerline{\epsfxsize 2.0truein\epsfbox{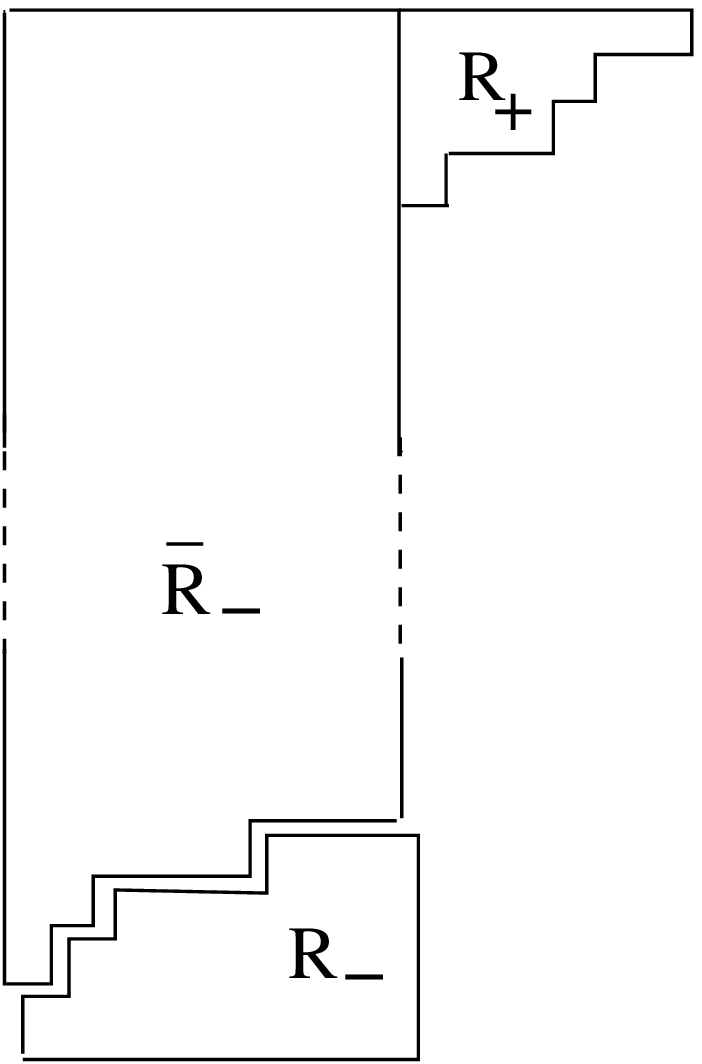}}
\leftskip 2pc
\rightskip 2pc \noindent{\ninepoint\sl
\baselineskip=2pt {\bf Fig.4}
{{The figure depicts an $SU(N)$ representation $R$ as a coupled representation $R_+{\bar R}_-$. The representation ${\bar R}_-$ is conjugate to representation $R_-$. }}}
\bigskip
Using this, and trading $N$ and $\theta$ for $t$ and $\bar t$ defined by
\kahlermoduli , the $q$YM partition function becomes
\eqn\large{\eqalign{
Z^{q{\rm YM}}(\Sigma_{g >1})
&=
\a(g_s,\theta)
\sum_{l\in {\bf Z}}\sum_{R_{+},R_{-}}\Bigl( S_{00}\ {\rm dim}_q R_{+}\bar R_{-}\Bigr)^{2-2g}
q^{\frac{p}{2}\left(\k_{R_{+}}+\k_{R_{-}}\right)}e^{i\theta\left(\vert
R_{+} \vert - \vert R_{-} \vert \right)} \cr
& ~~~~~~~\times q^{{p N \over 2}\left(\vert R_{+}
\vert +\vert R_{-} \vert \right)}
e^{-\frac{(t^2-{\bar t}^2)l}{2(p+2g-2) g_s} }
e^{-\frac{p(t+\bar t)l^2}{2(p-2+2g)}} e^{
-pg_sl\left( \vert R_{+}\vert -\vert R_{-} \vert\right) }.
}}
So far, we have merely rewritten \large\  and now we are ready to turn to its
factorization at large $N$.
The key fact, shown in Appendix B, is the following relation
\eqn\relii{\eqalign{
q^{\rho^2 +{N\over 24} }\ S_{00}\ {\rm dim}_q (R_{+} \bar R_{-})=&
{M(q)\ \eta^N(q) \over K_{R_{+}R_{-}}(Q,q)} W_{R_{+}}^2(q) W_{R_{-}}^2(q)\cr
&\times (-)^{\vert R_{+} \vert +\vert R_{-} \vert }
 q^{-{1\over 2}(\k_{R_{+}}+ \k_{R_{-}})}
Q^{-{1\over 2}(\vert R_{+} \vert +\vert R_{-})}}}
where $M(q)$ is MacMahon function, $\eta(q)=q^{1\over 24}\prod_{j=1}^{\infty} (1-q^j)$
and
\eqn\krs{
K_{R_{+}R_{-}}(Q,q):=\sum_P Q^{\vert P \vert}W_{PR_{+}}(q)W_{PR_{-}}(q)
}
Moreover, the normalization factor $\a(g_s,\theta),$ defined in \alpii,
can be written as
\eqn\norma{
\a(g_s,\theta) = \vert \Upsilon (t,g_s) \vert^2
\Biggl( q^{\rho^2+ {N\over 24} }\Biggr)^{2-2g}
}
where
\eqn\fgt{
 \Upsilon(t,g_s)=\exp\left(
-\frac{t^3}{6p(p+2g-2)g_s^2}+\frac{(p+2g-2) t}{24 p}\right).}
Using \relii\ and \norma\ we can recast
$Z^{q{\rm YM}}_{g>1}$ in the large $N$ limit as
a sum of ``chiral blocks'':
\eqn\decomp{
Z^{q{\rm YM}}(\Sigma_{g >1})=\sum_{ l\in {\bf Z}}\sum_{R_1,\ldots, R_{2g-2}}
Z^{q{\rm YM},+}_{R_1,\ldots, R_{2g-2}}\Bigl ( t + pg_s l\Bigr)
Z^{q{\rm YM},+}_{R_1,\ldots, R_{2g-2}}\Bigl ( {\bar t} - pg_s l\Bigr).
}
The chiral block $Z^{q{\rm YM},+}_{R_1,\ldots, R_{2g-2}}$
is defined by
\eqn\chir{\eqalign{
Z^{q{\rm YM},+}_{R_1,\ldots, R_{2g-2}}(t)=&{ Z}_0(t,g_s)\;\eta^{{t\, \chi  \over (p-\chi)g_s }}\
e^{-\frac{t\left( \vert R_1\vert
+\ldots \vert R_{2g-2}\vert \right) }{(p-2+2g)} }\cr
\times &\sum_R q^{{1\over 2}(p+2g-2)\k_R}e^{-t\vert R
\vert}{\ W_{R_1R}(q) \ldots W_{R_{2g-2}R}(q)\over W_{0R}(q)^{4g-4}}
}}
which agrees, as we will discuss below, with the perturbative topological string amplitudes
with $(2g-2)$ stacks of D-branes inserted in the fiber.  The extra
prefactor $\eta^{{t\, \chi  \over (p-\chi)g_s }} $ needs to be explained.
However this factor has only a genus 0 contribution perturbatively.
Namely, by using the modular property of the Dedekind eta function, we
find that this factor contributes as
$$\eta^{{t\, \chi  \over (p-\chi)g_s }}\sim {\rm exp}\left(-{ct\over g_s^2}\right)
+({\rm non-perturbative}),$$
where
$c={\pi^2 \chi\over 6(p-\chi)}$.  It is reassuring that this
can be viewed as a correction to the topological string
amplitudes at genus zero.  This is possible because the power of $t$
is less than three and the fact that topological string partition function
at genus $0$
is ambiguous up to the addition of a quadratic polynomial in $t$.  Thus
we can redefine the degree 0 contributions \pertii\ to be
\eqn\gammagt{
{\widehat Z}_0(t,g_s)=Z_0(t,g_s)\ {\rm exp}\left(-{ct\over g_s^2}\right). }

We will come back to the interpretation of the blocks in the
topological string context, after we have discussed the genus zero case.

\subsec{Factorization into Chiral Blocks for Genus $g=0$ Case.}
To find the large $N$ limit of the $q$-deformed
YM partition function on a genus zero Riemann surface
$Z^{q{\rm YM}}(S^2)$ we proceed in a similar way, except that we write the quantum dimension of the coupled
representation $R_{+} {\bar R}_{-}$ slightly differently:
(see Appendix B)
\eqn\reli{\eqalign{
q^{\rho^2+{N\over 24}} \ S_{00}\ {\rm dim}_q (R_{+} \bar R_{-})=
&M(q) \eta^N(q) N_{R_{+}R_{-}}(Q,q)(-)^{\vert R_{+} \vert +\vert R_{-} \vert }\cr
&\times  q^{-{\k_{R_{+}}+ \k_{R_{-}} \over 2}}
Q^{-{\vert R_{+} \vert +\vert R_{-} \vert \over 2}},
}}
where $q=e^{-g_s},Q=e^{-g_s N}$ as before, and
\eqn\nrs{
N_{R_{+}R_{-}}(Q,q):=\sum_P (-)^{\vert P \vert} Q^{\vert P \vert}W_{P
R_{+}}(q)W_{P^T R_{-}}(q).
}
Using \reli\ and \nrs\ as well as \seckas , we find the large $N$
limit
of
$Z^{q{\rm YM}}(S^2)$ given by \ymred\ is
\eqn\decompzero{
Z^{q{\rm YM}}(S^2)= \sum_{l\in {\bf Z}}\sum_{R_1,R_2}
Z^{q{\rm YM},+}_{R_1,R_2}\Bigl ( t + pg_sl\Bigr)
Z^{q{\rm YM},-}_{R_1,R_2}\Bigl ( {\bar t} - pg_sl\Bigr),
}
where the chiral block, $Z^{q{\rm YM},+}_{R_1,R_2}(t)$,
is defined by
\eqn\zplus{\eqalign{
Z^{q{\rm YM},+}_{R_1,R_2}(t)
={\widehat Z}_0(t,g_s)& \ q^{ \frac{\k_{R_1}+\k_{R_2} }{2} }
e^{-\frac{t\left( \vert R_1\vert + \vert R_2 \vert
\right) }{(p-2)} }\cr
\times &\sum_R q^{\frac{(p-2)\k_R}{2}}e^{-t\vert R \vert}\ W_{R_1R}(q)
 W_{R R_2^T}(q).}
}
and the `anti-chiral block', unlike the higher genus case, is transposed:
$$
Z^{q{\rm YM},-}_{R_1,R_2}({\bar t})=Z^+_{R_1^T,R_2^T}({\bar t})(-)^{\left( \vert R_1\vert + \vert R_2
\vert \right) }
$$

\subsec{Large N Expansion and Perturbative Topological String}

To summarize the results of the previous two
subsections, we have found that in the large $N$ expansion the
partition function of the $q$-deformed YM theory
corresponding to
$N$ D4-branes on the zero section of $L_1$ in
$$
X_{\Sigma} = L_1 \oplus L_2 \rightarrow \Sigma
$$
factorizes as
\eqn\factorizes{
Z^{q{\rm YM}}(\Sigma) = \sum_{ l \in {\bf Z}}\sum_{R_1,\ldots, R_{|2g-2|}}
Z^{q{\rm YM},+}_{R_1,\ldots R_{|2g-2|}}(t+ l p g_s)
Z^{q{\rm YM},+}_{R_1,\ldots R_{|2g-2|}}(\bar t- l p g_s),}
where $p$ is related to the degree of $L_1$ by ${\rm deg}(L_1)=-p$.
What is the interpretation of the chiral block in terms of the
topological string theory on $X$?

For concreteness, let us focus on $g\geq 1$ case. According to
\chir, the chiral blocks are expressed
as
$$
Z^{q{\rm YM},+}_{R_1,\ldots, R_{\vert 2g-2\vert }}(t)= {\widehat Z}_0
e^{-\frac{t\left( \vert R_1\vert
+\ldots \vert R_{2g-2}\vert \right) }{(p-2+2g)} }
\sum_R {W_{R_1R}(q) \ldots W_{R_{2g-2}R}(q)\over W_{0R}(q)^{4g-4}}q^{{1\over 2}(p+2g-2)\k_R}e^{-t\vert R
\vert},
$$
where ${\widehat Z}_0$ contains  the classical pieces  \gammagt.

First notice that in the large $N$ expansion of the $q$-deformed YM partition
function, there is a limit where the anti-chiral theory decouples, namely
taking $\bar t \rightarrow \infty$, at fixed $t$. Here we are treating $t$ and $\bar t$ as independent variables.
In this limit only the trivial representation contributes in the sum over
$R_1,...,R_{2g-2}$ in \decomp\  because
of the factors $e^{-\frac{\bar t\left( \vert R_1\vert
+\ldots \vert R_{2g-2}\vert \right) }{(p-2+2g)} }$ in $Z^{q{\rm YM}-}$.
Thus, we find
$$
\lim_{\bar t
\rightarrow \infty}Z^{q{\rm YM}}(t,\bar t) = Z^{q{\rm YM}, +}_{0,\ldots, 0}(t),
$$
up to a trivial anti-chiral piece.
A look back at section 2 shows that this is exactly equal
to the perturbative {\it closed} topological string partition function \geng\
for the same Calabi-Yau:
$$
Z^{q{\rm YM},+}_{0,\ldots,0}(t) = Z^{{\rm top}}(X_{\Sigma_g})(t).
$$
Thus, in the limit where the
anti-chiral part of the $q$YM theory decouples, we recover
the perturbative topological string theory amplitude on $X$,
at the value of K\"ahler parameter fixed by the attractor mechanism.

However, this is clearly not all, and  $|Z^{{\rm top}}(X_{\Sigma_g})(t)|^2$
is but the first in the sum over chiral blocks in the large $N$ expansion of
$q{\rm YM}$ on $\Sigma_g$. Amazingly, the other chiral blocks $also$ have an
interpretation in terms of theory on $X_g$, but now involving D-branes!
Another look back at section 2 shows that the object which appears in the higher chiral blocks is also topological string amplitude on $X$, but
with with $(2g-2)$ D-branes {\it in the fiber} over
$(2g-2)$ points on $\Sigma$ given in equation \infib ,
with the degrees of the line bundles adjusted to the Calabi-Yau at hand, $i.e.$
${\rm deg}(L_1)=2g-2+p,\;{\rm deg}(L_2) = -p$. Moreover, the D-branes are
moved off the Riemann surface by an amount $t/(p+2g-2)$.
With this understanding, we have the identity:
$$
Z^{q{\rm YM},+}_{R_1,\ldots, R_{\vert 2g-2\vert }}(t)=
Z^{{\rm top}}_{R_1,\ldots, R_{\vert 2g-2\vert }}(t),
$$
It should be clear from the preceding discussion that
$Z^{{\rm top}}_{R_1,\ldots, R_{\vert 2g-2\vert }}(t)$
can itself be obtained as a $\bar t \to \infty$ limit of the $q$YM
amplitude -- this time one obtained by
gluing $(4g-4)$ ``dual''
pant vertices \dualpant\ to get a Riemann surface of genus $g$ with
$(2g-2)$ punctures \ref\anv{M. Aganagic, A. Neitzke, C. Vafa, to appear}.

To understand the geometric meaning of the chiral blocks,
it is useful to express the large $N$-expansion of
$Z^{q{\rm YM}}$ in terms of an integral over $(2g-2)$ holonomies
$V_1,\cdots, V_{2g-2} \in U(\infty)$ as
\eqn\nice{\eqalign{
Z^{q{\rm YM}}(\Sigma_{g >1})
=\sum_{l\in {\bf Z}}
\int dV_1 \cdots dV_{2g-2}& \ Z^{{\rm top}}\Bigl(g_s, t+pg_s
l; V_1, \ldots,
V_{2g-2}\Bigr)
\cr
&\times Z^{{\rm top}}\Bigl(g_s, \bar t-pg_s
l; V_1^{-1}, \ldots,
V_{2g-2}^{-1}\Bigr)  ,}}
where
$$
Z^{{\rm top}}\Bigl(g_s, t;V_1, \ldots, V_{2g-2}\Bigr)
=
\sum_{R_1,\ldots, R_{2g-2}} Z^{{\rm top}}_{R_1,\ldots, R_{2g-2}}\Tr_{R_1}V_1
\ldots \Tr_{R_{2g-2}}V_{2g-2},
$$
and we used the orthogonality of the characters \orth .
Each of the D-branes on the fiber described in section 2.1
intersects with $C_4$ at a non-contractible circle on its worldvolume.
Thus, one can  regard $V_i$ ($i=1,...,2g-2$)
as a holonomy of the gauge field
on the $i$-th stack of D-branes around the cycle, keeping
track of the way the worldsheet ends on the D-branes. We take
the number of D-branes at each stack to be infinite so that
the representations $R_1,...,R_{2g-2}$ can be arbitrary.

These D-branes are directly related to the presence of
$\Omega$-points in the large $N$ limit of the ordinary 2d YM theory \gt .
To see the connection, it is useful to move to a more geometric
basis for the chiral blocks. We can do this by using the Frobenius
formula, which expresses the trace $\Tr_R(U)$ for any representation
$R$ in terms of a sum of products of traces of $U$ in
the fundamental representations as
\eqn\frobenius{ \Tr_R(U) = {1\over n!} \sum_{\sigma\in S_n} \chi_R(\sigma)
\tr_{\vec k(\sigma)}V,}
where
$$
\tr_{\vec k(\sigma)}V = \prod_{i=1} \left(\tr V^i\right)^{k_i(\sigma)}.$$
The sum in \frobenius\ is over elements of symmetric group $S_n$ of $n$
elements, where $n=|R|$ is the number of boxes in the Young diagram corresponding to the representation $R$,
$\chi_R(\sigma)$ is the character of the representation of $S_n$ corresponding
to the same Young diagram, $\tr$ is the trace over the fundamental
representation, and  $k_i(\sigma)$ is the number of cycles of
length $i$ in the permutation $\sigma \in S_n$. Note that
$\tr_{\vec k(\sigma)}V$ depends only on the conjugacy class of the permutation
$\sigma$.
By using the orthogonality relations \orth\ it follows from \frobenius\ ,
$$ \int dV \;\tr_{\vec{k}}V \; \tr_{\vec k'} V^{- 1}
 = \delta({\vec k}, {\vec k'}) \zeta({\vec k}), $$
where
$$\zeta({\vec k}) =\prod_{i=1}^\infty i^{k_i^a} k_i^a!.$$
One can re-express \nice\ as
\eqn\cycles{
Z^{q{\rm YM}}(\Sigma_g)  = \sum_{ l\in {\bf Z}}
\sum_{{\vec k}_1, ..., {\vec k}_{2g-2}}
 Z^{{\rm top}}_{{\vec k}^1,...,{\vec k}^{2g-2}}(g_s, t+ pg  l)
\; Z^{{\rm top}}_{{\vec k}^1,...,{\vec k}^{2g-2}}
(g_s, \bar t- pg  l)
\prod_{a=1}^{2g-2} \zeta({\vec k}^a)
,}
where $Z^{{\rm top}}_{{\vec k}^1,...{\vec k}^{2g-2}}$ is defined by
\eqn\topcycles{
Z^{{\rm top}}(g_s,t;V_1,...,V_{2g-2})= \sum_{{\vec k}^1,...,{\vec k}^{2g-2}}\;
Z^{{\rm top}}_{{\vec k}^1,...{\vec k}^{2g-2}} (g_s,t)\;
\tr_{\vec{k}^1} V_1 \cdots \tr_{\vec{k}^{2g-2}} V_{2g-2}.}

By construction, $Z^{{\rm top}}_{{\vec k}^1, ... , {\vec k}^{2g-2}}(g_s,t)$
is the topological string amplitude with the constraint that
the worldsheet ends on the $a$-th stack of D-branes with $k_i^a$ boundaries
wrapping on the non-contractible cycle $i$-times ($i=1,2,...$).
The expression \cycles\
suggests that the complete topological string partition
function includes gluing of the holomorphic and anti-holomorphic topological
string amplitudes together so that the boundaries of the holomorphic
and the anti-holomorphic worldsheets match up on the D-branes.
Note that the combinatorial factor $\prod_{a=1}^{2g-2} \zeta({\vec k}^a)$
is exactly the number
of ways the boundaries can be glued together.
Since $Z^{{\rm top}}$ is an exponential of a sum of connected worldsheets,
the full topological string partition function involves
arbitrary numbers of connected worldsheets, holomorphic and
anti-holomorphic. Thus, despite the appearance of D-branes,
worldsheets contributing to the topological string amplitudes
are still closed, except that they are piecewisely holomorphic
or anti-holomorphic.

In this way, the $2g-2$ stacks of D-branes generate analogues of the
$2g-2$ $\Omega$ points that appear in the large $N$ limit of the
ordinary $2d$ YM \gt.
To complete the story, one would need to explain
why exactly $2g-2$ D-branes are involved in the non-perturbative
completion of the string theory on
$L_1 \oplus L_2 \rightarrow \Sigma_g$. We will come back to this below.

So far we have focused on higher genus $g\geq 1$
Riemann surfaces. Things work in an analogous
way for $S^2$. In this case,
the chiral $q$YM
blocks given by
$$
Z^{q{\rm YM},+}_{R_1,R_2}=
{\widehat Z}_0 q^{ \frac{\k_{R_1}+\k_{R_2} }{2} }
e^{-\frac{t\left( \vert R_1\vert + \vert R_2 \vert
\right) }{(p-2)} }\sum_R q^{\frac{(p-2)\k_R}{2}}e^{-t\vert R \vert}\ W_{R_1R}(q)
 W_{R R_2^T}(q)
$$
are computing partition functions of topological strings with
two infinite stacks of D-branes in the fibers over 2 points on the $S^2$,
$$
Z^{q{\rm YM},+}_{R_1R_2}(t) = Z^{{\rm top}}_{R_1,R_2}(t),
$$
where the right hand side is computed using technology of section
2, up to the (ambiguous) pieces $Z^{{\rm top}}\rightarrow {\widehat Z}_0 Z^{{\rm top}}$
which for us are defined in \gammagt.
We can write the above in an alternative way which makes contact with the
topological vertex.
\eqn\topvert{
Z^{q{\rm YM},+}_{R_1R_2}(t) = {\widehat Z}_0 \ q^{ \frac{\k_{R_1}}{2} }\ e^{-{t\left( \vert R_1\vert +\vert R_2\vert \right)\over p-2}}
\sum_{R}
\
e^{-t\vert R \vert}\ q^{\frac {(p-1)\k_R}{2}} C_{0R_1R^T}(q)
C_{0RR_2}(q),
}
where $C_{R_1 R_2 R_3}$ is the topological vertex defined in \AKMV.
This is the partition function
of the topological A-model on ${\cal{O}}(-p)\oplus{\cal{O}}(p-2)\rightarrow {\bf P}^1$
with non-compact Lagrangian D-branes inserted at the two of the four external
lines in the web-diagram. Note that for $p=1$
the above expression for the topological string amplitude
is not completely satisfactory since
$$
t = {1\over 2} (p-2) N\, g_s\, -i\theta,
$$
so $Re(t)\leq 0$. The meaning of this in the black hole context
is that, even though we started with $N$ D-branes, their effective charge
is negative, and they really correspond to anti D-branes.
In terms of the topological string, we should rewrite the amplitude in terms of
the flopped geometry with the K\"ahler parameter given by
$$
\hat t = i \pi-t,
$$
with $Re ({\hat t}) >0.$
Indeed, as we show in appendix C, we can
rewrite the chiral block $Z^{q{\rm YM},+}_{R_1,R_2}(t)$ as follows:
\eqn\flop{
Z^{q{\rm YM},+}_{R_1,R_2}(t)={\widehat Z}_0(t)
(-)^{-{1 \over 12}}e^{-{{\hat t}\over 12}}
\sum_R (-)^{|R|} e^{-{\hat t}|R|}
C_{R_1 R_2 R}(q) C_{00R^T}(q).
}

Analogously to $g>1$ case, we can write partition sum as
\eqn\niceii{
Z^{q{\rm YM}}(S^2)=\sum_{l\in {\bf Z}}\int dV_1 dV_2 \
\; Z^{q{\rm YM},+}\Bigl(t+pg_sl; V_1,V_2\Bigr)\;
Z^{q{\rm YM},-}\Bigl({\bar t}-pg_sl;
 V_1^{-1},V_2^{-1}\Bigr),}
where
$$
Z^{q{\rm YM},+}\Bigl(t;V_1,V_2\Bigr)
=\sum_{R_1,R_2} Z^{q{\rm YM},+}_{R_1,R_2}(t)\; \Tr_{R_1}V_1\; \ \Tr_{R_2} V_2 .
$$

\subsec{Topological String Interpretation}
Above we have found that there is an apparent discrepancy
between the prediction of \osv\ that $Z_{BH}=|Z^{{\rm top}}|^2$ and
the explicit computation of the black hole ensemble which leads to
$$Z_{BH}=\sum_\alpha |Z_{\alpha}^{{\rm top}}|^2.$$
Moreover there is an extra sum over integers:  This extra sum has
been explained in \V\ as being related to summing over RR fluxes
in the geometry (or alternatively it is required for making the
chemical potential has suitable periodicity in the imaginary
direction).  However the extra sum over chiral blocks labelled by
$\alpha$ which is given by topological string amplitudes with
D-branes may appear to be in contradiction with \osv .  It turns
out that there is no contradiction \anv  , and this is related to
the fact that the Calabi-Yau is non-compact and has more moduli
coming from the non-compact directions.  Taking these into account
is equivalent to writing the black hole partition function in
terms of the D-brane blocks as given above.  See \anv\ for more
detail.

\newsec{The 4d Black Holes and the $c=1$ String.}

As is well known, the mirror symmetry implies that the B-model on the
conifold
\eqn\con{
zw - xy = {\hat t}
}
is the same as a $\hat t \rightarrow 0$ limit of the A-model topological string on
$O(-1)\oplus O(-1) \rightarrow {\bf P}^1$,
where $\hat t$ is the K\"ahler parameter of the ${\bf P}^1$.  On
the other hand, it is also well known that B-model topological string
on the conifold is the same as the bosonic $c=1$ string at the self dual
radius, where $\hat t$ is identified with the cosmological constant
\ghv.
Given the results of the previous sections,
we should also be able to give a non-perturbative formulation
for the B-model topological string on the conifold, and hence also of the
$c=1$ string theory.
In other words, it follows by the mirror symmetry that the small $\hat t$ limit of
$Z^{q{\rm YM}}(S^2)$ gives the non-perturbative formulation of the $c=1$ string.

Recall that perturbative $c=1$ amplitudes
depend effectively only on the ratio $\mu$
of the cosmological constant and $g_s$,
$$
\mu = {\hat{t}\over g_s} = {N\over 2}+ i\alpha,
$$
where $\a = (\pi -\theta)/g_s$. One would expect that this extends to
non-perturbative contributions as well
-- the only finite mass D-brane in the B-model on the conifold is a D3-brane wrapping the
$S^3$ of size $\hat{t}$, whose action is then $\mu={\hat{t}\over g_s}$.
Given this, taking a small $\hat{t}$ limit at fixed $\mu$, is equivalent to
taking to a small $g_s$ limit keeping $\mu$ fixed.

Consider the partition function of
the $q$YM for $g=0, p=1$ obtained in section 4,
\eqn\finformii{
Z^{q{\rm YM}}(S^2)=\left( {2\pi \over g_s}\right)^{N/2}\ q^{N\over 12}\
\ S_{00}(q,N) \
\sum_{n \in {\bf Z}^N} \exp\left( -{2\pi^2 n^2 \over g_s}+
\frac{2\pi}{g_s} \theta \sum_{i=1}^N n_i \right).
}
The small $g_s$ limit of the
$S_{00}(g_s,N)$ factor is well known to be (see \wd)
given by $\frac{(2\pi g_s)^{N^2\over 2}}{vol (U(N))}
\left( {2\pi \over g_s}\right)^{N/2}.$
Then, in the sum over instantons only the ${\vec n}=0$ sector survives in the limit, and we
have a prediction for the {\it non-perturbative} partition function $Z_{c=1}$
of the $c=1$ string as
\eqn\finformii{Z_{c=1}=
\frac{ (2\pi g_s)^{\frac{N^2}{2}} }
{{\rm vol}(U(N))}\
\left( {2\pi \over g_s}\right)^{N}.
}
%

The answer suggests what is the underlying theory which provides
the non-perturbative completion of $c=1$ string. It is simply the
2D topological YM theory. This can be seen by taking the $g_s
\rightarrow 0$ limit of the qYM path integral. This theory should
be the effective theory of $N$ D3 branes on the non-compact
3-cycle which is dual to the $S^3$ of the conifold (in the same
sense as 2d qYM is effectively describing the $N=4$ topological YM
on the D4 branes.).

In studying this theory at large $N$, we can use the results of the
large $N$ limit of $q$YM theory that we already studied which led to
$$ Z^{q{\rm YM}}(S^2)= \sum_{l\in {\bf Z}}\sum_{R_1,R_2} Z^{q{\rm
YM},+}_{R_1,R_2}\Bigl ( t + g_sl\Bigr) Z^{q{\rm
YM},-}_{R_1,R_2}\Bigl ( {\bar t} - g_sl\Bigr)
$$
and take the $g_s\rightarrow 0$ limit of that, which is the same as
giving a chiral decomposition of \finformii\ at large $N$.

Using the results in the appendix C,
it is easy to show that
\eqn\limch{
A_{R_1 R_2}=\lim_{g_s \to 0}Z_{R_1 R_2}
}
is given by
$$
A_{R_1,R_2}(\mu)=a(g_s,\mu){{\rm dim} R_1(\mu) \ {\rm dim} R_2^T
(\mu)\over {\rm vol}(U(\mu))}.
$$
Here ${\rm dim} R(\mu)$ is given by
$$
{\rm dim} R(\mu)=d(R) \prod_{\tableau{1}\in R}
(\mu-i(\tableau{1})+j(\tableau{1})),
$$
where $i,j(\tableau{1})$ label the location, $i.e.$ the row and the column,
of the box in the Young tableaux of $R$.
The coefficient $d(R)$ is related to the dimension of the corresponding symmetric group representation
$$
\quad d(R)=\prod_{\tableau{1}\in R}\frac{1}{h(\tableau{1})},
$$
where $h(\tableau{1})$ is the hook length of the corresponding box in $R.$
Finally,  coefficient $a(g_s,\mu)$
is given by
$$
a(g_s,\mu)=(2\pi g_s)^{\frac{\mu^2}{2} }\left( \frac{2\pi}{g_s} \right)^{\mu}
e^{\frac{i 2\pi \mu^2}{3}}e^{i\delta},
$$
where the phase $e^{i\delta}$ is independent of $\mu$ and
can be attributed to
the ambiguity of the definition of the chiral block.
By ${\rm vol}(U(\mu))$ we denote a function obtained by
expanding the volume of the unitary group $U(M)$, in powers of $1/M$,
and in the result, setting $M=\mu$.

Thus, we find that, at large $N$ the non-perturbative $c=1$
partition function factorizes as

\eqn\answ{Z_{c=1} \approx \sum_{l\in {\bf Z}}\sum_{R_1,R_2}
A_{R_1,R_2}\Bigl ( \mu -l\Bigr) A_{R_1^T,R_2^T}\Bigl ( {\bar \mu} +
l\Bigr) (-)^{( \vert R_1\vert +\vert R_2\vert)}}

The equality  holds $only$ in the asymptotic expansion, and we have
denoted this  by ``$\approx$''.

It is natural to ask what the interpretation of chiral blocks
$A_{R_1 R_2}$ is. First of all, note that the vacuum chiral block
$$
A_{00}(\mu) = {a(g_s,\mu) \over  {\rm vol}(U(\mu))},
$$
in the $1/\mu$ expansion is precisely the vacuum amplitude of the
$c=1$ string. The higher chiral blocks in \answ\ are related
to the scattering amplitudes of
perturbative $c=1$ string at self-dual radius \DMP.

Recall that $c=1$ string is free when written in terms of fermions.
Correspondingly, the S-matrix elements are diagonal in terms of fermions,
\eqn\ferm{
S_{R}(\mu)=\langle R\vert S \vert R\rangle ,
}
and can be brought into the form (see for example
\ADKMV ),\foot{{}From now on we will not be careful about
 constant factors like $\pi$,\ $2$ etc}
$$
S_R=\prod_{i=1}^{\#{\rm diag}(R)} \Gamma(i \mu + f_i +1)
\Gamma( -i\mu + h_i +1) \cos\left[{\pi \over 2}(f_i+i\mu)\right]
\cos\left[{\pi \over 2}(h_i-i\mu)\right],
$$
where $\#{\rm diag}(R)$ stands for the number of boxes
on the diagonal of $R$ and the free fermion state $\vert R \rangle$
is expressed as:
$$
\vert R \rangle=\prod_{j=1}^{\#{\rm diag}
(R)}\psi_{-(h_i+\half)}\psi^*_{-(f_i+\half)}\vert 0\rangle ,
$$
where the fermion and hole momenta are defined as
$f_i=R_i-i$ and $h_i=R^T_i-i$.
The similarity with our blocks becomes apparent if we use
$$
\Gamma(x) \Gamma(1-x)={\pi \over \sin(\pi x)}
$$
to rewrite $S_R$ as
$$
S_R=\prod_{i=1}^{\#{\rm diag}(R)}\frac{\Gamma( i\mu + f_i +1)}{
\Gamma(i \mu - h_i )}{\cos\left[{\pi \over 2}(f_i+i\mu)\right]
\cos\left[{\pi \over 2}(h_i-i\mu)\right]
\over \sin \left[\pi(i\mu-h_i)\right]}
$$
$$
=\exp\left[ -{i \pi \over 2}\sum_{i=1}^{\#{\rm diag}(R)}(h_i+f_i) \right]
\prod_{\tableau{1}\in R}(i\mu-i({\tableau{1}})+j(\tableau{1}))+O(e^{-\mu})
$$
Note that for $c=1$ string the cosmological constant $\mu$ is real, while
in the topological string context $\mu$ is complex parameter.
In what follows we'll take advantage of the
analytic continuation provided by the topological string, and
make no particular distinction between $\mu$ and $i \mu$.
With this in mind, in the expression we recognize elements of \answ ,
\eqn\scat{S_R(\mu)=\exp\left[- {i \pi \over 2}\sum_{i=1}^{\#{\rm diag}(R)}
(h_i+f_i) \right]
{{\rm dim} R (\mu) \over d(R)} .}

We will express $Z_{c=1}$ as an overlap of D-brane wave functions
below.  This will also allow us to interpret the result
as a tachyon scattering amplitude.

\subsec{$Z_{c=1}$ as an Overlap of D-Brane Wave Functions.}
It is natural to expect,
given the discussion of the preceding sections, that the
higher chiral blocks are computed by a particular
D-brane amplitude on the B-model conifold geometry.
In the $c=1$ string theory language, this would correspond to
a particular tachyon
scattering amplitude \ADKMV .

Moreover, we expect to be able to formulate
$Z_{c=1}$, in the large $N$ expansion as an overlap of D-brane wave functions
\eqn\overlp{
Z_{c=1} = \int {\cal D}x_1 {\cal D}x_2 |Z_{{\rm D-brane}}(x_1,x_2)|^2,
}
for some appropriate D-brane configuration and appropriate
measure ${\cal D}x$ on the space of hermitian matrices of infinite rank.
Using the relation of the B-model on the conifold with $c=1$ string,
$Z_{D-brane}$ can also be thought of as a particular
coherent state scattering amplitude of the $c=1$ string
$$
Z_{{\rm D-brane}}=Z_{{\rm scatt.}}.
$$
We will show that, $Z_{{\rm D-brane}}(x_1,x_2)$
is a wave function in holomorphic polarization given by
$$
Z_{{\rm D-brane}}(x_1,x_2) = \sum_{R_1,R_2} A_{R_1,R_2^T} \Tr_{R_1} x_1 \Tr_{R_2} x_2
$$
Moreover there is a natural measure
\eqn\measure{
{\cal D} x = dx d{\bar x} \; \exp\left[ \tr (x {\bar x})\right]
}
over commuting Hermitian matrices $x,\bar x$ with the property that
$$
\int {\cal D}x\; \Tr_{R_1} x \,\Tr_{R_2} x=\delta_{R_1,R_2},
$$
so that \overlp\ holds.

Consider $c=1$ amplitude with coherent state of incoming and outgoing tachyons turned on
\eqn\bos{
S(t,\bar t)= \langle t\vert S\vert \bar t\rangle ,
}
where
$$
\vert t\rangle = \exp\left(\sum_{n=1}^{\infty}{t_n\over n} \alpha_{-n}
\right)|0\rangle ,
$$
and $\alpha_{-n}$ are the usual boson creation operators corresponding
here to the tachyon momentum modes.
We can relate this to a D-brane correlation function in the B-model
on the conifold \con . As explained in \ADKMV , the effective B-model theory
is a theory of a chiral boson (corresponding to the tachyon)
on the Riemann surface\foot{View $x,\bar x$ here as independent complex variables.}
$$
x {\bar x} = \mu,
$$
and the fermions of this chiral boson are the D-branes.
The Riemann surface is a cylinder with
two asymptotic regions corresponding to large $x$ or large $\bar x$.
Sending outgoing and ingoing tachyon pulses is equivalent by bosonization
to placing D-branes at points
$x_i$, ${\bar x}_i$ near one of two boundaries on the Riemann surface.
The $x_i,\bar x_i$, viewed as
eigenvalues of infinite dimensional matrices $x$
and $\bar x$, are related to coherent states of tachyons
by
\eqn\eleven{
t_n =\, \tr x^n, \quad \bar{t}_n =\, \tr {\bar x}^n.
}
This provides an identification between $\vert t\rangle$ and
D-brane configurations
$\vert x\rangle$,
$$\vert t \rangle = \vert x\rangle.$$

The scattering amplitudes \bos\  are related
by bosonisation to the formula
\ferm\ we had above. Namely, bosonisation relates
$$
|R\rangle =
\sum_{\vec k}{ {\chi_{R}(C(\vec{k})) \over z_{\vec{k}}}}|\vec{k}\rangle ,
$$
where
$$
\vert \vec{k} \rangle =\prod_{n} (\alpha_{-n})^{k_n} \vert 0 \rangle .
$$
Note that in terms of $\langle x | R\rangle = \Tr_R x$ and
$\langle x|\vec k\rangle = \tr_{\vec k} x$ this is
just the relation \frobenius\ where $z_{\vec k}$ is also defined.

{}From this we can write $S(t,\bar t)$ as
$$
S(t,\bar t) = \sum_R S_R \; \Tr_R x \; \Tr_R {\bar x} ,
$$
where $S_R$ is the scattering amplitude in \ferm . The left hand side is better thought of in terms of tachyon scattering, the right in terms of D-brane (or fermion) amplitudes. More generally, any tachyon scattering amplitude can be related to a D-brane correlation function, by bosonisation.

We will now see that
$$\sum_{R_1,R_2} A_{R_1,R_2^T}\Tr_{R_1} x_1 \Tr_{R_2} x_2$$
does correspond to a scattering amplitude of the $c=1$ string, and thus
to a D-brane amplitude.
First of all, using properties of tensor product coefficients
$N_{R_1 R_2}^R$,\foot{These satisfy
$\sum_{R} N_{R_1 R_2 }^{R} \Tr_{R} x = \Tr_{R_1} x \; \Tr_{R_2} x$
essentially by definition, and relation we need
$\sum_{R_1,R_2} N_{R_1 R_2}^{Q} {\rm dim}(Q) = {\rm dim}(R_1){\rm dim}(R_2)$
is a special case of this when $x$ is the identity matrix, $x = id$.}
and the definition of $S_Q(\mu)$ in \scat\ it is easy to see that, up
to a $\mu$-independent phase (and a trivial prefactor), the above is equal
$$
\sum_{R_1,R_2} \; \sum_{Q}
N_{R_1 R_2}^{Q} \;d(Q)\;S_{Q} \;\Tr_{R_1} {x}_1 \;\Tr_{R_2} {x}_2
$$
This can be directly interpreted as a tachyon scattering amplitude as follows. First it is easy to prove\foot{For example, evaluate the above formula by contracting with arbitrary $\langle x \vert$ and prove that $
\exp(\sum_n {1\over n} \tr x^n\; \tr {\bar x}^n) = \sum_R \;\Tr_R x \;\Tr_R {\bar x}
$.}
that
$$
\sum_{R_1,R_2,R} N_{R_1R_2}^R \Tr_{R_1} {x}_1 \;\Tr_{R_2} {x}_2 \vert R \rangle =
\vert t_1+t_2 \rangle
$$
where $\vert t_{1,2}\rangle$ are coherent states corresponding to $\vert x_1\rangle$
and $\vert x_2\rangle$, and we defined
$$\vert t_1 + t_2 \rangle = \exp\Bigl(\sum_{n} {1\over n} (t_{1,n} + t_{2,n}) \alpha_{-n} \Bigr)\vert 0 \rangle$$
Finally, using
$$
{1\over n!}\;(\alpha_{-1})^{n} \vert 0 \rangle=
\sum_{R}\; {1\over n!}\;\chi_{R}(1^{n}) \vert R \rangle
=\sum_{R}\; d(R)\;\vert R\rangle
$$
(the sum is over representations $R$ of $n$ boxes), we see that
$$
\sum_{R_1,R_2} A_{R_1,R_2^T} \Tr_{R_1} x_1 \Tr_{R_2} x_2
=\langle 1\vert S \vert {t}_1+{t}_2\rangle
$$
where
$$\vert 1 \rangle \; :=\exp(\alpha_{-1})\vert 0\rangle.$$
To summarize, we found that
$
\sum_{R_1,R_2} A_{R_1,R_2^T} \Tr_{R_1} x_1 \Tr_{R_2} x_2
$ corresponds to a tachyon scattering amplitude, and thus also to a D-brane correlation function.
To complete the claim that this in fact gives the
D-brane or scattering amplitude which computes $Z_{c=1}$,
we need a natural inner product where the overlap of the
above wave functions would agree with the $1\over \mu$ expansion of
$Z_{c=1}$.

In the context of tachyon scattering this is straightforward.
Recall that there is a canonical measure on the
space of coherent states $\psi(t)=\langle t \vert \psi \rangle$
$$
\langle \psi\vert \chi \rangle =\int {\cal D}t \;\bar{\psi}(\bar t) \;\chi(t)
$$
where
$$
{\cal D}t = \prod_{n=1}^{\infty}
{1\over n} dt_n \,d\bar{t}_{n}\; \exp\left(-{t_n{\bar t}_n\over n}\right).
$$
This is the same inner product as the natural inner product in the
$\vert\vec{k}\rangle$ or $\vert R\rangle$ basis --
in other words, $\int {\cal D}t \;\vert t \rangle \langle {\bar t}\vert$
is the identity operator\foot{For example, one can easily check that
$\int {\cal D}t \;\langle {\vec k}\vert \,t\,\rangle \langle\, {\bar t}\,\vert\,{\vec k}'\,\rangle = {z_{\vec{k}}}\; \delta(\vec{k},\vec {k}')
$ as it should be to agree with the definition of $\vert{\vec k}\rangle$.}.
Putting everything together, it immediately follows
\eqn\bhe{Z_{c=1} = \int {\cal D}t_1{\cal D} t_2 \; \vert Z_{{\rm scatt.}}(t_1,t_2)\vert^2
}
where
$$
Z_{{\rm scatt.}}(t_1,t_2)=\langle 1\vert S \vert {t}_1+{t}_2\rangle .
$$
The above expression is very reminiscent of the idea of \KazakovPM :  The
state $\langle 1|$ above is a chiral version of a black hole state.  It is
as if in the non-perturbative formulation of the $c=1$ theory we have been
forced to consider a black hole state.  Moreover the formula \bhe\ is analogous
to computing the decay of the black hole state to arbitrary tachyon state.  It would
be very interesting to develop this picture further, especially in view
of the fact that we have an exact non-perturbative formulation in terms of which
\bhe\ is only an asymptotic large charge expansion.

We could stop here, however, this way of writing $Z_{c=1}$ is
not very natural if one wants to relate it D-branes.
To do this we need an inner product
in the $x$-basis corresponding to D-brane positions.
$$
\langle \psi\vert \chi \rangle =\int {\cal D}x \;\bar{\psi}(\bar x)\; \chi(x)
$$
We can use define this by asking that it is compatible with the
inner product in the $R-basis$, i.e.
$$
\int {\cal D}x \;\Tr_{R} x \;\Tr_Q {\bar x} = \langle Q\vert \R \rangle =
\delta_{Q,R}
$$
We will show in the appendix D, that the inner product exists, and
can explicitly be given in terms of \measure\ as claimed above.

We have thus shown that we can alternatively write $Z_{c=1}$
in terms of D-brane amplitude
$$
Z_{{\rm D-brane}}(x_1,x_2) = \sum_{R_1,R_2} A_{R_1,R_2^T} \Tr_{R_1} x_1 \Tr_{R_2} x_2
$$
as
$$
Z_{c=1} = \int {\cal D}x_1{\cal D} x_2 \;\vert Z_{{\rm D-brane}}(x_1,x_2)\vert^2,
$$
and where
$$
Z_{{\rm D-brane}}(x_1,x_2)=Z_{{\rm scatt.}}(t_1,t_2)
$$
with $t_{1,2}$ and $x_{1,2}$ related by \eleven .

\bigskip
\medskip

\centerline{\bf Acknowledgments}

We are grateful to C. Beasley, J. Bryan, M. Marino, A. Neitzke and
R. Pandharipande W. Taylor and E. Witten for valuable discussions.  H.O. and C.V.
thank the 2004 Simons Workshop on mathematics and physics, for stimulating
talks.
H.O. in addition thanks the Aspen Center for Physics and the High Energy
Theory Group at Harvard University for their hospitalities.

 The research of M.A. was
supported in part by a DOE OJI Award, and an Alfred P. Sloan
Foundation fellowship.
The research of H.O. was supported in part
by DOE grant DE-FG03-92-ER40701.
The research of N.S. and C.V. was supported
in part by NSF grants PHY-0244821 and DMS-0244464.

\vfill
\eject

\appendix{A}{The Cap Amplitude of 2d ($q$)YM and the Fourier Transform}

In this appendix we compute the path integral on a disk of the
ordinary 2d YM theory and its $q$-deformed version,
in $\Phi$ basis instead of the usual holonomy basis.
In the holonomy basis, it is given by:
$$
Z_{{\rm 2dYM}}(C)(U) = \sum_\R {\rm dim}(\R) \Tr_\R U,
$$
where as before $U = Pe^{\oint A}$.
The Fourier transform to the $\Phi$ basis is given by the following path
integral over the boundary of the disk,
$$
Z_{{\rm 2dYM}}(C)(\Phi) = \int
dU \;e^{{1\over g_s} \oint_{\partial C}\Tr \Phi A}\; Z_{{\rm 2dYM}}(C)(U).
$$
Since the YM path integral localizes to configurations where $\Phi$ is covariantly constant,so in particular $\Phi$ and $A$ commute, integrating over the angles gives (see \MM\ for details. There, effectively the same matrix integrals were considered in a related context)
$$
Z_{{\rm 2dYM}}(C)(\Phi) = \int \; \prod_i du_i\;
\Delta_H(u) \;e^{{1\over g_s}\sum_i{{\vec{\phi}\cdot{\vec u}}}}\;
Z_{{\rm 2dYM}}(C)({\vec u}),
$$
where we defined an anti-hermitian matrix $u$ by
$ U = e^{u},$
and
$
\Delta_H(u)= \prod_{\alpha >0}\sin({\vec\alpha} \cdot {\vec u}).
$
comes from the Haar measure on $U(N)$.
We can compute the integral by using
$$
\Tr_\R U := \chi_\R(\vec{u})= {\sum_{\omega \in w} (-1)^{\omega}
e^{\omega({\vec \lambda_\R}+{\vec \rho})\cdot {\vec u}}
\over {\sum_{\omega \in w} (-1)^{\omega}
e^{\omega({\vec \rho})\cdot {\vec u}}}},
$$
where $\lambda_\R$ is the highest weight vector of the
representation $\R$ and $\vec{\rho}$ is the Weyl vector;
the Weyl denominator formula
$$
\prod_{\alpha>0} \sin({\vec\alpha}\cdot{\vec u})=
{\sum_{\omega \in W} (-1)^{\omega} e^{\omega({\vec \rho})\cdot {\vec u}}};
$$
and by writing the dimension of representation $R$ as
\eqn\df{
{\rm dim}(\R) = \lim_{t\rightarrow 0}\chi_\R(t\vec{\rho}).
}
Plugging this into the integral, and performing a sum over the weight lattice
we get
$$
Z_{{\rm 2dYM}}(C)(\phi) = \lim_{t\rightarrow 0}\int \prod_i du_i
\;e^{\sum_i{{\vec{\phi}\cdot{\vec u}}}/g_s}\;\;
{\sum_{\omega \in W} (-1)^{\omega} \delta({\vec u} -t{\omega(\vec{\rho})})
\over\prod_{\alpha>0} \sin(t{\vec \alpha}\cdot{\vec\rho})},
$$
or, by computing the integral
$$\eqalign{
Z_{{\rm 2dYM}}(C)(\phi) &= \lim_{t\rightarrow 0}
{\sum_{\omega \in W} (-1)^{\omega} e^{t \vec{\phi}\cdot {\omega(\vec{\rho})}}
\over\prod_{\alpha>0} \sin({t\vec{\alpha} \cdot\vec\rho})}\cr
&= \lim_{t\rightarrow 0}{
{\prod_{\alpha>0} \sin({t\vec \alpha\cdot \vec\phi})}
\over
\prod_{\alpha>0} \sin({t\vec \alpha\cdot \vec\rho})}.}
$$
Finally, this is equal to the expression we gave in section 3
$$
Z_{{\rm 2dYM}}(C)(\phi) = \prod_{i<j} {\phi_i - \phi_j\over i-j},
$$
up to a constant multiplicative factor $\prod_{i<j} (i-j)$ which
we had dropped there.

Note that the analogue of this for the $q$-deformed amplitude,
$$
Z_{q{\rm YM}} = \sum_\R {\rm dim}_q(\R) \Tr_\R U ,
$$
is simply obtained by setting
$$
t= g_s
$$
in the formula for the dimension of representation \df, and {\it not} taking the
small $t$ limit:
$$
{\rm dim}_q(\R) = \chi_\R(g_s \vec{\rho}).
$$
This gives the path integral on the disk for the $q$YM theory
$$
Z_{{\rm qYM}}(C)(\phi) = \prod_{i<j} {[\phi_i/g_s - \phi_j/g_s]_q\over [i-j]_q}
$$
as claimed in section 3 (where we needed the inverse of this
Fourier transform).

\appendix{B}{${\rm dim}_q (R{\bar S})$ in terms of $K_{RS}$ or $N_{RS}$}

The quantum dimension of the coupled representation has the form
\eqn\coupl{
{\rm dim}_q (R\bar S)= {\rm dim}_q R \ {\rm dim}_q S \prod_{i=1}^{c_R}
\prod_{j=1}^{c_S} {[S_j+R_i+N+1-j-i][N+1-j-i] \over
[S_j+N+1-i-j][R_i+N+1-j-i] },
}
where $c_R$ is the number of rows
in $R$ and $R_i$ is the number of boxes in the i-th row. Now we
let $q=e^{-g_s}$ and express ${\rm dim}_q R$ as
\eqn\quant{
{\rm dim}_q R=Q^{-{\vert R
\vert \over 2}}W_R(q^{-1}) \prod_{i=1}^{c_R}
\prod_{j=1}^{R_i}(1-q^{-i+j}Q), \quad Q=e^{-g_sN}.
}
We write each of the products in \coupl and \quant in the exponential form:
$$
\prod_{i=1}^{c_R} \prod_{j=1}^{c_S}\left(1-q^{S_j+R_i-j-i+1}Q\right)
=\exp\left(- \sum_{n=1}^{\infty} { g_1(q^n) Q^n \over n} \right),
$$
where $g_1(q)=\sum_{i=1}^{c_R} \sum_{j=1}^{c_S} q^{S_j+R_i-j-i+1}.$
$$
\prod_{i=1}^{c_R} \prod_{j=1}^{c_S}\left(1-q^{S_j-j-i+1}Q\right)
=\exp\left(- \sum_{n=1}^{\infty} { g_2(q^n) Q^n \over n} \right),
$$
where $g_2(q)=\sum_{i=1}^{c_R} \sum_{j=1}^{c_S} q^{S_j-j-i+1}.$
$$
\prod_{i=1}^{c_R} \prod_{j=1}^{c_S}\left(1-q^{R_i-j-i+1}Q\right)
=\exp\left(- \sum_{n=1}^{\infty} { g_3(q^n) Q^n \over n} \right),$$
where $g_3(q)=\sum_{i=1}^{c_R} \sum_{j=1}^{c_S} q^{R_i-j-i+1}.$
$$
\prod_{i=1}^{c_R} \prod_{j=1}^{c_S}\left(1-q^{-j-i+1}Q\right)
=\exp\left(- \sum_{n=1}^{\infty} { g_4(q^n) Q^n \over n} \right),
$$
where $g_4(q)=\sum_{i=1}^{c_R} \sum_{j=1}^{c_S} q^{-j-i+1}.$
$$
\prod_{i=1}^{c_R} \prod_{j=1}^{R_i}(1-q^{-i+j}Q)
=\exp\left(- \sum_{n=1}^{\infty} { f_R(q^n) Q^n \over n} \right),
$$
where $f_R(q)=\sum_{i=1}^{c_R} \sum_{j=1}^{R_i} q^{-i+j}.$

So that ${\rm dim}_q ( R\bar S)$ is re-casted as
\eqn\newqu{
{\rm dim}_q ( R\bar S)=Q^{-{\vert R \vert +\vert S \vert \over
2}}W_R(q^{-1}) W_S(q^{-1}) \exp\left( -\sum_{n=1}^{\infty}
{M_{RS}(q^n) Q^n \over n} \right),
}
where $M_{RS}(q)=g_1(q)+g_4(q)-g_2(q)-g_3(q)+f_R(q)+f_S(q).$

It turns out  that $M_{RS}(q)=f_{RS}(q)$
where
$$ f_{RS}(q)=(q-2+q^{-1})f_R(q)f_S(q)+f_{R}(q) +f_{S}(q).$$

Now, we relate ${\rm dim}_q (R \bar S)$ with the functions $K_{RS}(q)$
and $N_{RS}(q)$  defined as follows:
\eqn\krsii{ K_{RS}(Q,q):=\sum_P
Q^{\vert P \vert}W_{PR}(q)W_{PS}(q)= W_R(q)W_S(q)\prod_{i=1,
j=1}^{\infty} (1-Qx_i y_j)^{-1} ,} \eqn\nrs{ N_{RS}(Q,q):=\sum_P
(-)^{\vert P \vert} Q^{\vert P \vert}W_{P R}(q)W_{P^T S}(q)=
W_R(q)W_S(q)\prod_{i=1, j=1}^{\infty} (1-Qx_i y_j), } where
$x_i=q^{R_i-i+1/2},\quad y_j=q^{S_j-j+1/2}. $ We used the
definition of $W_{R_1 R_2}$ in terms of Schur functions $s_R$:
\eqn\wversus{ W_{P R}(q)=s_{R}\left(q^{-i+1/2} \right)
s_{P}\left(q^{R_i-i+1/2} \right) } and the properties of Schur
functions \eqn\prop{
\sum_{R}s_R(x)s_R(y)=\prod_{i,j}(1-x_iy_j)^{-1},\quad
\sum_{R}s_R(x)s_{R^T}(y)=\prod_{i,j}(1+x_iy_j). }

As follows from \IK\
the functions $K_{RS}$ and $N_{RS}$ are expressed in terms of $f_{RS}(q)$ as
\eqn\newkrs{
K_{RS}(Q,q)=K_{..}(Q,q)\ W_R(q) \ W_S(q) \ e^{\sum_{n=1}^{\infty} {f_{RS}(q^n)Q^n \over n}}
,}
\eqn\newnrs{
N_{RS}(Q,q)=N_{..}(Q,q) \ W_R(q) \ W_S(q) \ e^{-\sum_{n=1}^{\infty}{f_{RS}(q^n)Q^n \over n} }
,}
where we denoted the trivial representation with $R=.$ and where
$ N_{..}(Q,q)=(K_{..}(Q,q))^{-1}.$

Now using
$W_{R}(q^{-1})=(-)^{\vert R\vert}W_{R^T}(q)=(-)^{\vert R\vert}q^{-{\k_R\over 2}}W_R(q)$
we find the relations
\eqn\apeni{
{\rm dim}_q (R \bar S)= K_{..}(Q,q)N_{RS}(Q,q)(-)^{\vert R \vert +\vert S \vert }
 q^{-{\k_R+ \k_S \over 2}} Q^{-{\vert R \vert +\vert S \vert \over 2}},
}
\eqn\apenii{
{\rm dim}_q (R \bar S)= {K_{..}(Q,q)\over K_{RS}(Q,q)} W_R^2(q) W_S^2(q)(-)^{\vert R \vert +\vert S \vert }
 q^{-{\k_R+ \k_S \over 2}}  Q^{-{\vert R \vert +\vert S \vert \over 2}}.
}
Finally, we use that
$$ q^{\rho^2+{N\over 24}} S_{00}=M(q)\eta^N(q) \ N_{..}(Q,q)$$
to obtain the relations used in section 5:
$$
q^{\rho^2+{N\over 24} } S_{00}\ {\rm dim}_q (R \bar S)=M(q)\eta^N(q) \ N_{RS}(Q,q)(-)^{\vert R \vert +\vert S \vert }
 q^{-{\k_R+ \k_S \over 2}} Q^{-{\vert R \vert +\vert S \vert \over 2}},$$
$$ q^{\rho^2+{N\over 24}} S_{00}\ {\rm dim}_q (R \bar S)= {M(q)\eta^N(q) \over K_{RS}(Q,q)} W_R^2(q) W_S^2(q)(-)^{\vert R \vert +\vert S \vert }
 q^{-{\k_R+ \k_S \over 2}}  Q^{-{\vert R \vert +\vert S \vert \over 2}}.
$$

\appendix{C}{Expressing $p=1, g=0$ Chiral Blocks in terms of S-Matrix.}

Below we compute the genus g=0, p=1 chiral block in terms of S-matrix.
Let us  define $t_K$ as $t=i\pi +t_K$ and write the chiral block as
$$ Z^{q{\rm YM},+}_{R_1,R_2}(t)={\widehat Z}_0
e^{t\left(\vert R_1\vert+\vert R_2\vert\right)}q^{\k_{R_1}\over 2}
Z'_{R_1 R_2}(t_K),$$
where
$$Z'_{R_1 R_2}(t_K):= \sum_R (-)^{\vert R\vert} e^{-t_K \vert R\vert}C_{0R_1 R^T}(q)
C_{0RR_2}(q).$$

First, we use the definition of the vertex in terms of Schur functions
$$ C_{R_1 R_2 R_3}=q^{\k_{R_1} \over 2}s_{R_3}(q^{\rho})
\sum_{\eta} s_{R_1^T/ \eta}(q^{R_3 +\rho})s_{R_2/ \eta}(q^{R_3^T +\rho})
$$
to recast the sum $Z'_{R_1 R_2}(t_K)$ as
$$Z'_{R_1 R_2}(t_K)=(-)^{\vert R_2\vert}s_{R_1}(q^{\rho})
 \sum_R  s_{R}(\lambda q^{\rho +R_1})
 \sum_{\eta} s_{R/ \eta}(q^{-\rho}) s_{R_2^T / \eta}(q^{-\rho}),$$
where $\rho^i=-i+\half, \quad i=1, \ldots, \infty$ and $\lambda=e^{-t_K}.$

Next, we use the identities
$$ \sum_{R} s_{R}(\lambda q^{\rho +R_1})s_{R/ \eta}(q^{-\rho})=
s_{\eta}(\lambda q^{\rho +R_1}) \prod_{i,j}\left(1-\lambda q^{R_1^i-i+j}\right)^{-1}
$$
and
$$ \sum_{\eta}s_{R_2^T / \eta}(q^{-\rho})s_{\eta}(\lambda q^{\rho +R_1})=
\lambda ^{\vert R_2\vert} s_{R_2^T}(\lambda^{-1}q^{-\rho},q^{\rho +R_1})
$$
and bring $Z'_{R_1 R_2}(t_K)$ into the form
$$
Z'_{R_1 R_2}(t_K)=(-)^{\vert R_2\vert}s_{R_1}(q^{\rho})
\prod_{i,j}\left(1-\lambda q^{R_1^i-i+j}\right)^{-1}
\lambda^{\vert R_2\vert} s_{R_2^T}(\lambda^{-1}q^{-\rho},q^{\rho +R_1}).
$$
This can be further simplified  by using
$$ \prod_{i,j}\left(1-\lambda q^{R_1^i-i+j}\right)^{-1}=
\prod_{i,j}\left(1-\lambda q^{-i+j}\right)^{-1} \prod_{i=1}^{c_{R_1}}
\prod_{j=1}^{R_1^i}\left(1-\lambda q^{-i+j}\right) .$$
Now we recall the formula for quantum dimension
$$ {\rm dim}_q R_1 (q,\lambda)=\lambda^{- {\vert R_1 \vert \over 2}} s_{R_1}(q^{-\rho})
\prod_{i=1}^{c_{R_1}}
\prod_{j=1}^{R_1^i}\left(1-\lambda q^{-i+j}\right),$$
and compare $Z'_{R_1 R_2}(t_K)$ with $W_{R_1 R_2^T}(q,\lambda):={S_{R_1 R_2^T}\over S_{00}}$
$$ W_{R_1 R_2^T}(q,\lambda)={\rm dim}_q R_1(q,\lambda) \lambda^{\vert R_2\vert \over 2}
s_{R_2^T}(\lambda^{-1}q^{-\rho},q^{\rho +R_1}).
$$
We use $s_{R_1}(q^{\rho})=(-)^{\vert R_1 \vert} q^{-{\k_{R_1} \over 2}}
s_{R_1}(q^{-\rho})$ to find
$$ Z'_{R_1 R_2}(t_K)=(-)^{\vert R_1 \vert + \vert R_2\vert}e^{-\frac{t_K}{2} \left( {\vert R_1 \vert + \vert R_2\vert} \right)}q^{-{\k_{R_1} \over 2}}
N_{..}(\lambda,q)W_{R_1 R_2^T}(q,\lambda),$$
where $\lambda=e^{-t_K},\quad
N_{..}(\lambda,q)=\prod_{i,j}\left(1-\lambda q^{-i+j}\right)^{-1}.$
The chiral block is then expressed in terms of $W_{R_1 R_2^T}$ as
follows
\eqn\main{
Z^{q{\rm YM},+}_{R_1,R_2}(t)={\widehat Z}_0(t)\ N_{..}(\lambda,q)
e^{\frac{t_K}{2} \left( {\vert R_1 \vert + \vert R_2\vert} \right)}
W_{R_1 R_2^T}(q,\lambda),}
where $t=i\pi+t_K.$

If we now let $\hat t=-t_K$ and express $W_{R_1 R_2^T}(q,\lambda)$ as in \IKii
$$W_{R_1 R_2^T}(q,\lambda)= {e^{\frac{\hat t}{2} \left( {\vert R_1 \vert + \vert R_2\vert} \right)}\over N(e^{-\hat t},q)}
\sum_R (-)^{|R|} e^{-{\hat t}|R|} C_{R_1 R_2 R}(q) C_{00R^T}(q),
$$
we find the ``flopped'' expression for the chiral block:
\eqn\flopapp{
Z^{q{\rm YM},+}_{R_1,R_2}(t)={\widehat Z}_0(t)
\frac{N(e^{\hat t},q)}{N(e^{-\hat t},q)}
\sum_R (-)^{|R|} e^{-{\hat t}|R|}
C_{R_1 R_2 R}(q) C_{00R^T}(q).
}

The final step is to use the relation \IKii
$$ N(Q,q)=N(Q^{-1},q)(-Q)^{-{1\over 12}}$$
to bring the flopped chiral block into the form:
\eqn\finappc{
Z^{q{\rm YM},+}_{R_1,R_2}(t)={\widehat Z}_0(t) (-e^{\hat t})^{-{1\over 12}}
\sum_R (-)^{|R|} e^{-{\hat t}|R|}
C_{R_1 R_2 R}(q) C_{00R^T}(q).
}

\appendix{D}{The Inner Product of D-Brane Wave Functions in $c=1$ String}
Here we show that the matrix integral
$$
\int {\cal D} x \;\Tr_R x \;\Tr_Q {\bar x}=\int d x d \bar x\;
\exp\left[\tr (x {\bar x})\right]
\;\Tr_R x \;\Tr_Q {\bar x}.
$$
in the definition of the $c=1$ string overlap equals
$$\int {\cal D} x \;\Tr_R x \;\Tr_Q {\bar x}=
\delta_{Q,R}.
$$
This is a so called
``normal'' matrix integral, meaning that $x$ and $\bar x$
are commuting matrices.

We will begin with an analogous finite
$M$ integral and than take $M$ to infinity.
Integrating over the angles in the above formula is standard, where one gets
$$
{1\over M!} \int \prod_i \;dx_i\; dy_i\;
 \;\Delta(x)\; \Delta(\bar x)\;\exp\left[\tr (x \bar x)\right] \;\Tr_R x\; \Tr_Q \bar x
$$
It is useful here to change variables to $x$ and $z$ where
$$\bar x=z/x,$$ which gives
$$
{1\over M!}\int\prod_i\; {dx_i d z_i\over x_i}\; e^{\sum_i z_i}\;
\sum_{\wo,\wo'}(-1)^{\wo+\wo'}
\;\prod_{i}{x_i}^{R_{\wo(i)} - \wo(i)-Q_{\wo'(i)}+\wo'(i)}
{z_i}^{Q_{\wo'(i)} + M - \wo'(i)}
$$
where we have in addition used the trace formula
$$Tr_R A = {\det_{i,j}(A_i^{R_j+M-j})\over \Delta(A)},$$
which holds for any matrix $A$,
and written
$$
\det(A_{ij})=\sum_{\wo} (-1)^{\wo}  A_{1\wo(1)}\ldots A_{M\wo(M)}
$$
Integrating over $x_i$ gives zero, unless $R=Q$ and $\wo=\wo'$
and we are left with computing
$$
\prod_{i}\;\int dz_i \;e^{z_i}\;{z_i}^{Q_{i} + M - i} = \prod_{i=1}^{M}\;
(Q_i + M - i)!
$$
All in all this gives, for rank $M$ matrix
$$
{1\over {\rm vol}(U(M))}\int dx  d\bar x \;\exp\left[\tr (x \bar x)\right]
\; \Tr_R x\; \Tr_Q \bar x \propto
{(2\pi)^{{M^2\over 2}+{M\over 2}}\over {\rm vol}(U(M))}
{{\rm dim}(Q)\over d(Q)} \delta_{R,Q}.
$$
Taking $M\rightarrow \infty$ limit of this corresponding to matrices of infinite rank gives
$$
\lim_{M\rightarrow \infty}\int d x d {\bar x} \;
\exp\left[\tr (x \bar x)\right]\; \Tr_R x\; \Tr_Q \bar x =
\delta_{R,Q},
$$
where we used that, for large $M$ the dimension of $SU(M)$ representation $dim(R)$
becomes the dimension of the corresponding
symmetric group representation $d(R)$ (up to an infinite factor $M^{\vert R \vert}$ which we
absorb in $x$, $\bar x$ and factors such as $\pi^M$ which go into renormalizing the measure).
\listrefs
\end